\documentclass[journal]{IEEEtran}

\usepackage{hyperref}
\newcommand{\figref}[2][{}]{\hyperref[#2]{\figurename~\ref{#2}#1}} 

\ifCLASSINFOpdf
   \usepackage[pdftex]{graphicx}
\else
\fi

\usepackage{amsmath}
\usepackage{cleveref}
\usepackage[noadjust]{cite}
\usepackage{booktabs}
\usepackage{longtable}

\usepackage{listings}
\usepackage{courier}

\usepackage{tikz}
\def\checkmark{\tikz\fill[scale=0.3](0,.35) -- (.25,0) -- (1,.7) -- (.25,.15) -- cycle;} 

\begin{document}
\title{Hybrid Symbolic-Numeric Framework for Power System Modeling and Analysis}

\author{Hantao~Cui,~\IEEEmembership{Member,~IEEE,}
        Fangxing~Li,~\IEEEmembership{Fellow,~IEEE,}
        Kevin~Tomsovic,~\IEEEmembership{Fellow,~IEEE}
\thanks{H. Cui, F. Li, and K. Tomsovic are with the Department
of Electrical Engineering and Computer Science, The University of Tennessee, Knoxville,
TN, 37996 USA. E-mail: fli6@utk.edu.} %
\thanks{This work was supported in part by the Engineering Research Center Program of the National Science Foundation and the Department of Energy under NSF Award Number EEC-1041877 and the CURENT Industry Partnership Program.}
}

\markboth{Preprint to be submitted to IEEE Transactions on Power Systems}%
{Cui \MakeLowercase{\textit{et al.}}: Hybrid Symbolic-Numeric}

\maketitle

\begin{abstract}
With the recent proliferation of open-source packages for computing, power system differential-algebraic equation (DAE) modeling and simulation are being revisited to reduce the programming efforts. 
Existing open-source tools require manual efforts to develop code for numerical equations, sparse Jacobians, and discontinuous components.
This paper proposes a hybrid symbolic-numeric framework, exemplified by an open-source Python-based library ANDES, which consists of a symbolic layer for descriptive modeling and a numeric layer for vector-based numerical computation.
This method enables the implementation of DAE models by mixing and matching modeling components, through which models are described.
In the framework, a rich set of discontinuous components and standard transfer function blocks are provided besides essential modeling elements for rapid modeling.
ANDES can automatically generate robust and fast numerical simulation code, as well as and high-quality documentation.
Case studies present a) two implementations of turbine governor model TGOV1, b) power flow computation time break down for MATPOWER systems, c) validation of time-domain simulation with commercial software using three test systems with a variety of models, and d) the full eigenvalue analysis for Kundur's system.
Validation shows that ANDES closely matches the commercial tool DSATools for power flow, time-domain simulation, and eigenvalue analysis.

\end{abstract}

\begin{IEEEkeywords}
Power systems, open-source, DAE modeling, symbolic calculation, time-domain simulation.
\end{IEEEkeywords}

\IEEEpeerreviewmaketitle

\section{Introduction}
\IEEEPARstart{P}{ower} system modeling and transient simulation is a widely studied yet challenging topic. 
Digital computer-based simulation has been dominating in the industry and academia with both closed-source tools 
\cite{Jalili-Marandi2013}
and open-source tools 
\cite{chow1992toolbox,zhou1996object,milano2005open,Milano2013,cole2011matdyn,Zhou2017,Top2016,Cui2018e, cui2019cyber, li2020large}
widely used. 
Although simulation software comes with a set of built-in models, users will likely need to customize models for new devices or control algorithms. 

To develop new models for simulation software is to implement the model equations in a program that can interact with the predefined software architecture.
In general, there are two approaches to implement user-defined models (UDMs): programmatically or through a graphical user interface (GUI) \cite{Siemens,PowerTech}, which is usually not available in open-source tools due to complexity and lack of return.
Still, open-source tools are crucial for scientific research, but they require programming proficiency to develop new models on top of a deep understanding of the tool \cite{Milano2010}. 

Two advanced UDM solutions exist in open-source tools: Dome cards \cite{Milano2013} and the Function Mockup Unit (FMU) support in GridDyn \cite{Top2016}.
Dome cards are plain-text files containing model descriptions in the card protocol. Using a symbolic library under the hood, Dome uses cards to generate intermediate code that can be modified into final models. Although cards are flexible, they do not live with the simulation code, and manual tweaks are often required. 
On the other hand, FMU is compiled directly from Modelica, an equation-based modeling language. 
Modelica libraries such as OpenIPSL \cite{Baudette2018} have been developed for power system simulation.
Although FMU has excellent speed and interoperability through the Functional Mockup Interface (FMI),
it has seen few adoptions in power system tools due to path dependence\footnote{Most of the widely used commercial tools today have a vast library of built-in models, which started to accumulate long before FMU was invented.} and, technically, data structure\footnote{
In Modelica/FMU, models are written separately and used combinatorially. Some implementations even require the precompilation of all possible combinations.}
.

This work proposes a hybrid symbolic-numeric method aiming to reduce
the efforts for modeling differential-algebraic equation (DAE) 
in power systems while maintaining numerical performance with the help of a symbolic toolbox.
The proposed method can be applied to major programming languages. 
An implementation has been open-sourced as the ANDES library written in Python, a scripting language suitable for power systems research and rapid prototyping.
Different from Dome cards, symbolically defined models are part of
the library and distributed with the program. 
Main contributions are as follows:
\begin{enumerate}
    \item The proposed hybrid symbolic-numeric method allows simple
scripting of DAE models with descriptive equation strings
instead of hard-coded implementations.

\item ANDES is the first open-source power system tool that enables writing models from block diagrams using modular discontinuous components and modeling blocks (such as transfer functions and proportional-integral controllers).

\item The library can generate efficient and robust numerical code from
descriptive models for fast simulation.

\item By preserving numerical interfaces, it 
can accommodate models that are much easier to implement in the traditional numerical way.
\end{enumerate}

The prior works on symbolic modeling and our advancements are discussed in the following.
Decades ago, symbolic approaches to power flow modeling \cite{Alvarado1988}, optimization \cite{Dzafic2002,Dzafic2004}, and device transients modeling \cite{Alvarado1991,Alvarado1996,Gao2004} were introduced. 
The pioneering works well proved the concept but exposed a remaining issue: scalability.
Namely, symbolic equations must be written for each device instance rather than each model type \cite{Alvarado1996}.
For large systems, a massive number of repetitive symbolic equations need to be created,
which are difficult to maintain and solve.
Besides, any system topology change requires manual modification to equations and is thus prone to errors.
In contrast, the proposed library models the abstract model type in the symbolic layer, agnostic to test systems.
Therefore, the computation time to process symbolic equations scales to the number (and the complexity) of model types, not the number of devices in any particular test case. 
In the generated code, vectorization is utilized for speed,
thus equations of all devices of the same type are updated in the same function calls.

This paper is organized as follows. \Cref{sec:philosophy} discusses the motivations and design philosophy of the work.
\Cref{sec:symbolic} and \Cref{sec:numeric_layer} explain the techniques for the symbolic and numeric layers with sufficient examples. 
\Cref{sec:case-studies} presents case studies, including two implementations of the TGOV1 model, power flow for MATPOWER systems \cite{Zimmerman2011b}, time-domain simulation verification with DSATools TSAT using three test systems with a variety of models, and full eigenvalue analysis.
\Cref{sec:conclusions} concludes the proposed work.   

\section{Motivations and Design Philosophy}
\label{sec:philosophy}
The overarching goal of the proposed hybrid symbolic-numeric method and its implementation in ANDES for power system modeling and analysis is to \textit{make modeling as simple as describing equations and make simulations as fast as using crafted code}.
Simplifying DAE modeling renders the library easy to use and modify for research and education. 
Maintaining a fast simulation speed makes the library capable of running large-scale studies.
As discussed, a purely symbolic approach will not scale to large systems, and a purely numerical approach will not reduce the programming efforts.
Therefore, a hybrid approach is proposed to take advantage of symbolic and numeric approaches in one library.

The design philosophy is two-fold: 1) to enable descriptive modeling using provided modeling elements and blocks, and 2) enable robust and fast numerical simulation through code generation and vectorization.
The first item can be realized in the symbolic layer in which model developers can mix and match parameters, variables, discrete components to describe DAE models. 
The second item can be realized through code generation from symbolically defined equations and coordination of the numerical functions.

\begin{figure}[!t]
\centering
\includegraphics[width=\columnwidth]{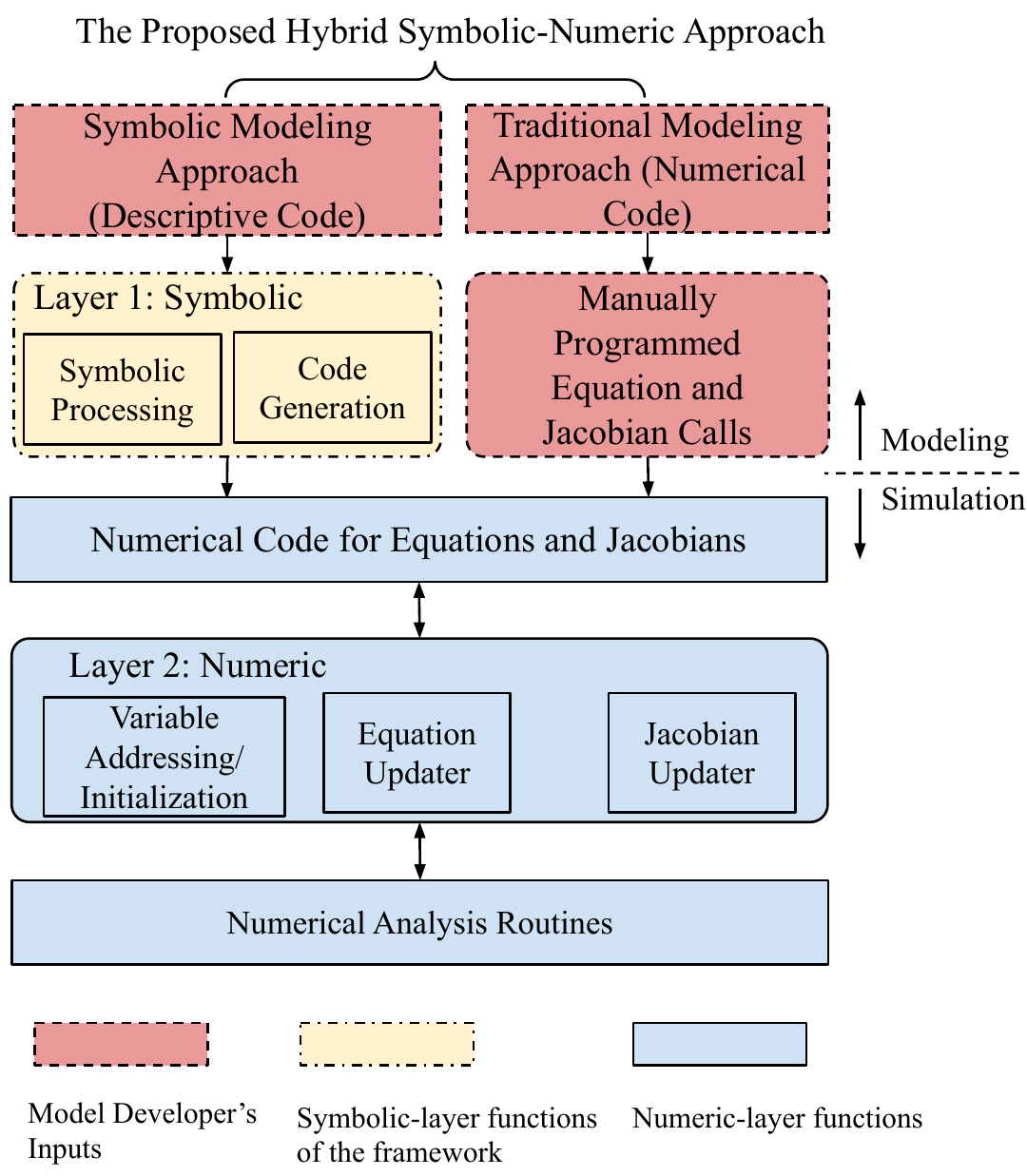}
\caption{Overview of the hybrid symbolic-numeric approach for modeling and simulation. Red boxes with dotted border indicate the required manual efforts.}
\label{fig:architecture_symbolic_library}
\end{figure}

\figref{fig:architecture_symbolic_library} shows the overview of the proposed hybrid symbolic-numeric framework with the upper part for hybrid modeling and the lower part for numerical simulation. 
This framework can accommodate two modeling approaches: 1) the proposed symbolic modeling approach using descriptive code, and 2) the traditional numerical modeling approach.
The symbolic approach is recommended due to simplicity and robustness because less programming is needed.
The symbolic layer can automatically generate symbolic equations and Jacobians, which, altogether, will be generated into loadable numerical code \cite{Meurer2017}. 
It ensures the same models will be used for simulation and documentation to achieve consistency between description and simulation.
Alternatively, the traditional numerical modeling approach can be used if a model cannot be easily implemented in the symbolic modeling approach.

The lower part of \figref{fig:architecture_symbolic_library} shows the numeric layer in the proposed framework for simulation. This layer organizes numerical code for equations and Jacobians, which include these generated by the symbolic layer and the manually written ones, to provide interface methods for addressing, initialization, and equation evaluation. 
Power system cases are loaded, and vector operations are utilized for optimal performance in a scripting environment.
Routine developers can develop specific numerical routines by calling the provided interface methods in specific orders.

The two-layer hybrid architecture also benefits the end-users who are not looking to develop models but instead use the library as a simulation tool.
Procedures in the symbolic layer only need to be executed once by the end-user, and the generated code will be serialized to disk for future reuse.
In terms of simulation performance, the proposed framework is on par with pure numerical libraries, since all computations in the numerical layer use vector operations.

\section{Symbolic Modeling Framework}
\label{sec:symbolic}
This section describes the implementation of the symbolic layer for the proposed library.
The symbolic layer covers class-based declarative modeling, symbolic processing, code generation, and automated documentation. 
Methods discussed in this section are exemplified in the Python language with the SymPy library but can be extended to other environments. 

\subsection{Basic Modeling Elements}
\label{sec-basic-elements}
The proposed library starts by observing that all DAE models can be described with a few categories of basic modeling elements.
Such categories include parameters, variables, discrete components, and services:

\begin{enumerate}
\item \textit{Parameters} are typically externally supplied data for defining specific devices.
\item \textit{Variables} either differential or algebraic, are the unknowns to be solved in the DAE system. Each variable is associated with values and an equation.
\item \textit{Discrete Components} describe the discontinuities, such as limits, associated with variables.
\item \textit{Services} are assisting types for simplifying expressions or fulfilling supplementary actions.
\end{enumerate}

The framework provides the above categories of modeling elements that can be instantiated to describe DAE models.
Modeling elements are containers in both symbolic and numeric layers.
In the symbolic layer, modeling elements are containers for metadata, such as name, description, unit, and equation strings.
In the numeric layer, they provide storage for associated numerical data, such as values and addresses.

\subsection{Classes for Descriptive Modeling}
\label{sec-class}
Python classes are the top-level containers to describe models.
A class for a DAE model can be created by defining class member attributes using the provided modeling elements.
The idea is best explained with a simple example, such as a constant shunt capacitor model for power flow given by
\begin{equation}
\label{eq:shunt_equations}
  \begin{array}{l}
    p_h = -gv_h^2 \\
    q_h = bv_h^2
  \end{array}
\end{equation}
where $h$ is the connected bus index, $v$ is the bus voltage, $p$ and $q$ are the power injections, and $g$ and $b$ are the conductance and susceptance, respectively. 
The implementation for the Shunt model is given in \autoref{lst:shunt_model} with the following remarks:
\begin{lstlisting}[float,floatplacement=H,caption={Shunt model for power flow (imports are omitted for simplicity).},captionpos=b,label=lst:shunt_model]
class Shunt(Model):
  def __init__(self):
    self.bus = IdxParam(info='bus index')
    self.g = NumParam(info='conductance', unit='pu')
    self.b = NumParam(info='susceptance', unit='pu')
    self.a = ExtAlgeb(model='Bus', indexer=self.bus,
                      src='a', e_str='g*v*v')
    self.v = ExtAlgeb(model='Bus', indexer=self.bus,
                      src='v', e_str='-b*v*v')
\end{lstlisting}

\begin{enumerate}
\item Lines 3-5 declares parameters \lstinline{bus}, \lstinline{g}, and \lstinline{b} for bus index, shunt conductance, and susceptance value.
\item Lines 6-9 declares external algebraic variables \lstinline{a} and \lstinline{v} for voltage phase and magnitude at the buses whose indices are \lstinline{bus}.
\item Lines 7 and 9 declares the active power load ($v^2 g$) and reactive power load  ($-v^2 b$) on the power balance equations associated with \lstinline{a} and \lstinline{v}.
\item The \lstinline{e_str} equation strings contain \lstinline{a}, \lstinline{v}, \lstinline{b} and \lstinline{b} strings, which are declared data attributes of the class.
\end{enumerate}

It is important to note that \autoref{lst:shunt_model} is an abstract Shunt model rather than just one particular Shunt device.
The Shunt model will host all Shunt devices of the same kind through vectorization so that only one invocation is needed for each equation.
An excellent discussion on this design choice can be found in Chapter 9.2 of \cite{Milano2010}.

Like a compiler, the underlying symbolic library requires a list of symbols to process equation strings. 
The base class \lstinline{Model} handles the bookkeeping of member attributes for all derived models.
Models can automatically capture the names and attributes instances to the corresponding storage in the declaration sequence based on attribute type. In Python, this is achieved by overloading the \lstinline{__setattr__()} protocol, which is invoked every time an attribute is assigned.
Therefore, the captured names will be converted to symbols for equation processing.
The approach allows us to keep the class definition concise while automatically performs the bookkeeping.

Therefore, the efforts to develop DAE models have been reduced.
All that required is to set up correct element containers and describe the mathematical equations.

\subsection{Discrete Components}
Discrete components such as limiters and deadlocks are common in practical models but are intricate to implement.
They often require manipulating equations and Jacobian, which, if not implemented correctly, can halt simulations.
In existing tools, discrete components are implemented \textit{ad hoc} and require manual efforts to be ported from one model to another.

The proposed library provides discrete components that are readily usable for describing DAE models.
Discrete components can export binary flags, which are evaluated in the numerical layer, to indicate the discontinuous status.
Flags can be used in equations to construct piece-wise equations with the benefit of not manipulating Jacobian matrices since discrete flags are preserved as variables in the corresponding derivative equations. 

\begin{figure}[!t]
\centering
\includegraphics[width=0.7\columnwidth]{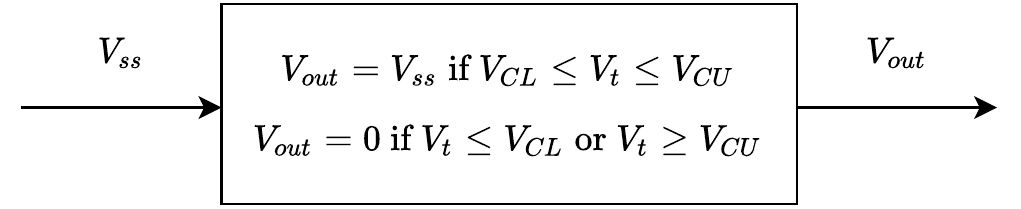}
\caption{A typical PSS final output limiter.}
\label{fig:pss-output-block}
\end{figure}

For example, a hard limiter takes an algebraic variable and two limit parameters as inputs and exports three flags, \lstinline{zi}, \lstinline{zl}, and \lstinline{zu} to indicate within, at the lower, and at the upper limits.
As a use case, consider a typical power system stabilizer (PSS) output limiter shown in \Cref{fig:pss-output-block}, where 
the final output $V_{out}$ depends on the terminal voltage $V_t$ and the given limits $V_{CL}$ and $V_{CU}$.
The output limiter can be conveniently implemented as in \autoref{lst:pss_output_limiter},
where Lines 1-2 creates a hard limiter called \lstinline{OL} that exports flags \lstinline{OL_zi},
\lstinline{OL_zl} and
\lstinline{OL_zu}.
Line 3 utilizes \lstinline{OL_zi} to construct the output variable \lstinline{Vout} with its equation through \lstinline{e_str} and the initial value equation through \lstinline{v_str}.
\begin{lstlisting}[float,floatplacement=H,caption={Stabilizer output limiter implementation.},captionpos=b,label=lst:pss_output_limiter]
self.OL = HardLimit(u=self.Vt, lower=self.VCL,
                    upper=self.VCU)
self.Vout = Algeb(e_str='vss*OL_zi-Vout',
                  v_str='vss*OL_zi')
\end{lstlisting}
\subsection{Services}
While the descriptive equation modeling is robust and straightforward, one needs to realize the limitation: vector operations are limited to arithmetic calculations. 
Descriptive equations cannot handle programmatic operations such as conditions and loops.
Services are helper types to overcome such limitations by allowing computing and storing values outside the DAE system using user-defined functions.
They are custom-computed but used in the same way as variables.

\begin{figure}[!t]
\centering
\includegraphics[trim={0cm 11cm 13cm 0},clip,width=\columnwidth]{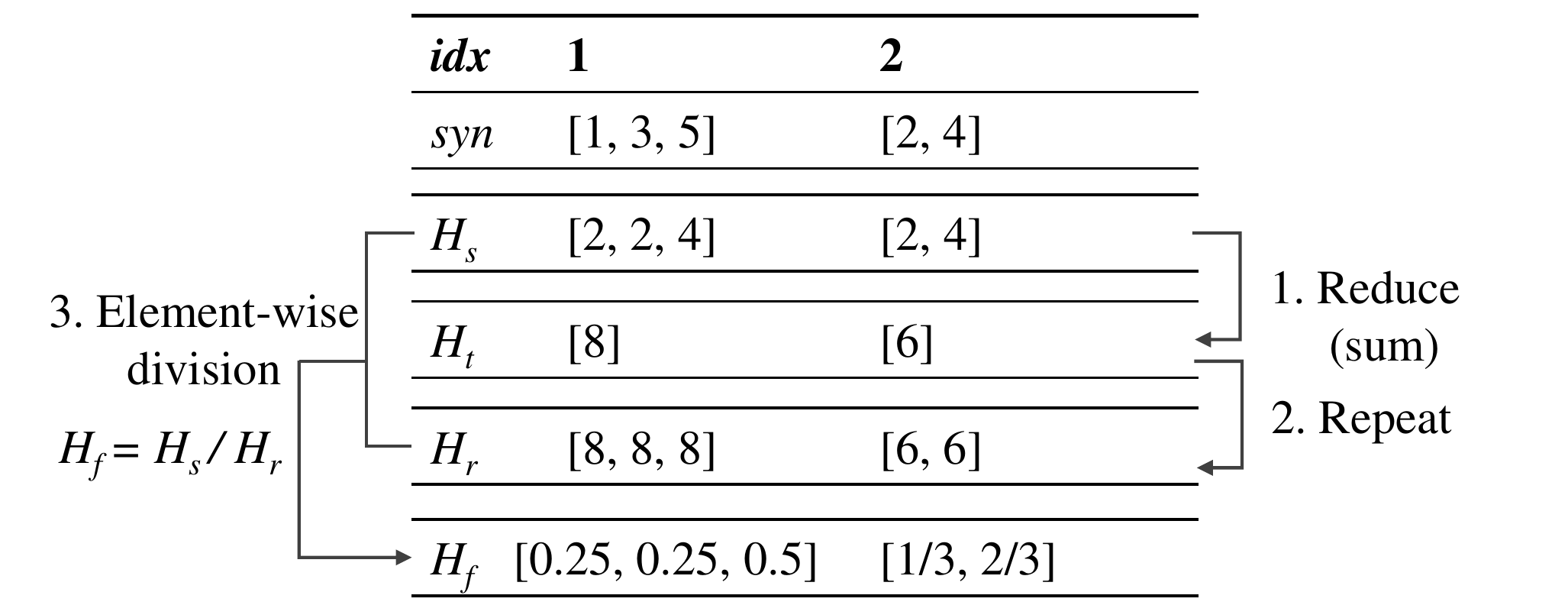}
\caption{Illustration of Reduce and Repeat services for COI.}
\label{fig:coi}
\end{figure}

An illustrative example is the calculation of inertia weights in the center-of-inertia (COI) model.
As shown in \Cref{fig:coi}, each COI device links to a number of generators (stored in \lstinline{syn}), retrieves their inertia $H_s$, and needs to compute the weights on the rotor speed for each linked generator.
A numerical program can quickly sum up the inertia and divide each inertia by the sum. 
However, since element-wise vector operations do not allow summation, the proposed library introduces two service types, one to reduce $H_s$ into $H_t$ using a summation function and the other to repeat the sum $H_t$ into the same shape as $H_s$. The element-wise division $H_s / H_t$ can be performed thereafter. 
\subsection{Modeling Blocks}
In addition to descriptive equation modeling, the library allows us to write models directly from transfer function diagrams.
A similar concept was reported in the InterPSS controller modeling language (CML) \cite{zhou_2012}, which utilizes the Java Annotation feature to provide a scripting environment for controller prototyping.
ANDES allows the composition of modeling elements into reusable modeling blocks, which can exports variables with equation templates.
Modeling blocks are instantiated as class member attributes like variables.
Upon instantiation, variable name placeholders in equation strings will be substituted with the actual names.
About 20 commonly used proportional-integral controllers and transfer functions, some with limiters, have been implemented.

\begin{figure}[!t]
\centering
\includegraphics[width=0.65\columnwidth]{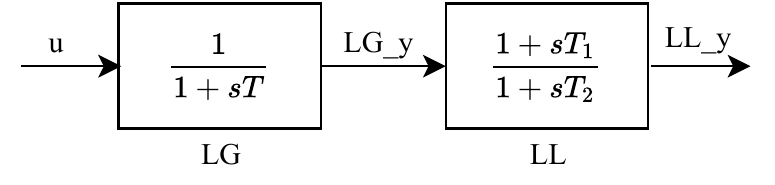}
\caption{A chained transfer functions example.}
\label{fig:chained-lag-lead-lag}
\end{figure}

For example, the chained transfer functions in \Cref{fig:chained-lag-lead-lag} can be implemented in barely two lines of self-explanatory code, as given by \autoref{lst:lag-lead-lag-block}.
Internally, model elements with their equations tailored with the instance name will be exported and captured by the hosting model class.
Block outputs are always named the block instance name with an underscore and letter \lstinline{y}.
In this example, Line 1 exports a differential variable named \lstinline{LG_y}, which is passed as an input in Line 2.
Similarly, the output of the lead-lag instance is accessible as \lstinline{LL_y}.

Modeling blocks can save efforts to reimplement equations in different models and improve readability.
In the meantime, modeling blocks can be mixed with custom descriptive equations using the exported variables whenever flexibility is needed.
\begin{lstlisting}[float,floatplacement=H,caption={Implementation of chained lag and lead-lag transfer functions.},captionpos=b,label=lst:lag-lead-lag-block]
self.LG = Lag(u=self.u, T=self.T, K=self.K)
Self.LL = LeadLag(u=self.LG_y, T1=self.T1, 
                  T2=self.T2)
\end{lstlisting}
\subsection{Symbolic Processing and Code Generation}
The symbolic processor converts the metadata, namely, equation strings, into symbolic expressions for symbolic Jacobian derivation, code generation, and documentation. 
These functionalities are part of the base \lstinline{Model} class and will be inherited by all derived models.
An external symbolic library is utilized to generate the symbolic expressions and Jacobian matrices for each model
with the following steps:

\begin{enumerate}{}
\item Prepare all variable symbols into a vector $\textbf{xy}$ in the declaration order so that each variable has a stable index. 

\item Convert each equation string to a symbolic expression (using \lstinline{sympy.sympify}).

\item Group differential and algebraic expressions into two vectors, $\textbf{f}$ and $\textbf{g}$, respectively, in the declaration order. 

\item Derive the expression vectors with respect to the ordered variable vector to obtain Jacobian matrices $\frac{d\textbf{f}}{d\textbf{xy}}$ and $\frac{d\textbf{g}}{d\textbf{xy}}$ (using \lstinline{sympy.Matrix.jacobian}).

\item Convert the Jacobian matrices to sparse to obtain non-zero triplets (\lstinline{row}, \lstinline{column}, \lstinline{value}), where \lstinline{row} is the index of the equation in the equation array, \lstinline{column} the variable index, and \lstinline{value} the derivative expression. 
\end{enumerate}

The following performance characteristics are relevant.
Symbolic processing is executed over each model, and thus the processing time scales linearly to the number of models. 
Each model only has a few to tens of equations; thus, the processing time is fast.
The processing is done before loading any system and is test-case independent. 

The symbolic processing for \lstinline{Shunt} is illustrated in \Cref{eq:shunt_xy,eq:shunt_g,eq:shunt_dg}. 
The Jacobian derivation and triplet conversion shown in \Cref{eq:shunt_dg} are automated with the symbolic library.
\begin{equation}
    \label{eq:shunt_xy}
    \textbf{xy} = [a, v]
\end{equation}
\begin{equation}
    \label{eq:shunt_g}
    \textbf{g} = [v^2g, -v^2b]
\end{equation}
\begin{equation}
    \label{eq:shunt_dg}
    \frac{d\textbf{g}}{d\textbf{xy}} = 
    \underbrace{
    \begin{bmatrix}{}
    0 & 2vg \\
    0 & -2vb
    \end{bmatrix}}_\text{dense}
    \xrightarrow{}
    \underbrace{
    \begin{bmatrix}
    (0, 1, 2vg) \\
    (1, 1, -2vb)
    \end{bmatrix}}_\text{sparse (row, column, value)}
\end{equation}

Code generation generates and stores numerical functions that are executable and will return the values of expressions. 
The code generation feature of the external symbolic library is utilized in the following steps:
\begin{enumerate}
    \item Generate numerical functions for each initialization, differential and algebraic equations, and each element in Jacobian Matrices (using \lstinline{sympy.lambdify}).
    \item For each Jacobian matrix, store the equation index \lstinline{row}, variable index \lstinline{column}, and the anonymous function for \lstinline{value} correspondingly in lists.
\end{enumerate}

It is important to note that \lstinline{row} and \lstinline{column} are local to each model and only depends on the number of declared variables. 
The following remarks are relevant. 
\begin{enumerate}{}
    \item In terms of performance, the generated numerical functions use the efficient NumPy library for vectorial computation and thus runs as fast as manually crafted code. 
    \item The overhead for symbolic processing and code generation can be eliminated by reusing the generated code through efficient serialization and de-serialization. 
    \item The library also takes manually written numerical function calls, as long as indices are provided and functions have the same signature as the generated code. This feature can be helpful to reuse existing numerical code.
\end{enumerate}

At this point, executable numerical code is obtained from the symbolically described DAE models. 

\subsection{Documentation}
Code documentation is essential for disseminating open-source research but is often underappreciated. The situation is understandable because maintaining documentation can take as much as, if not more than, the development efforts. All the existing power system simulators rely on manual efforts to document the implemented models.

The proposed library can automatically document the implemented equations for DAE models developed using declarative classes. 
Human-friend equations can be generated from symbolic expressions by substituting in \LaTeX-formatted variable strings. 
The documentation feature completes the symbolic layer to ensure the same models are used for simulation and documentation. 
To the best knowledge of the authors, the proposed library is the first in power system tools capable of generating equation documentation directly from source code.
For interested readers, the documentation is available online \cite{Cui2020rtd}, and the model documentation is under Section ``Model References". 

\section{Numeric layer Implementation}
The numeric layer establishes data structure for vector operations, and dispatches generated numerical code for the procedures in numerical simulation, such as setting initial values,  updating equations, and building Jacobian matrices.

\label{sec:numeric_layer}

\subsection{Data Structure and Vector Storage}
The numeric layer relies on arrays and sparse matrices to properly store data associated with declared elements.
Numerical values belonging to a modeling element instance are stored in the instance attributes.
Depending on the type, an element instance may contain member attributes for addresses, values, and equations. \Cref{tab:types_numerical_attributes} shows the supported attributes of the element types. 
Each address, value, and equation value attributes are stored as an array with its length equal to the number of devices. 
For example, if a particular system contains three \lstinline{Shunt} devices, attributes \lstinline{b} and \lstinline{g} will each contain a value array \lstinline{v} with a length of three.

\begin{figure}[!t]
\centering
\includegraphics[width=0.8\columnwidth]{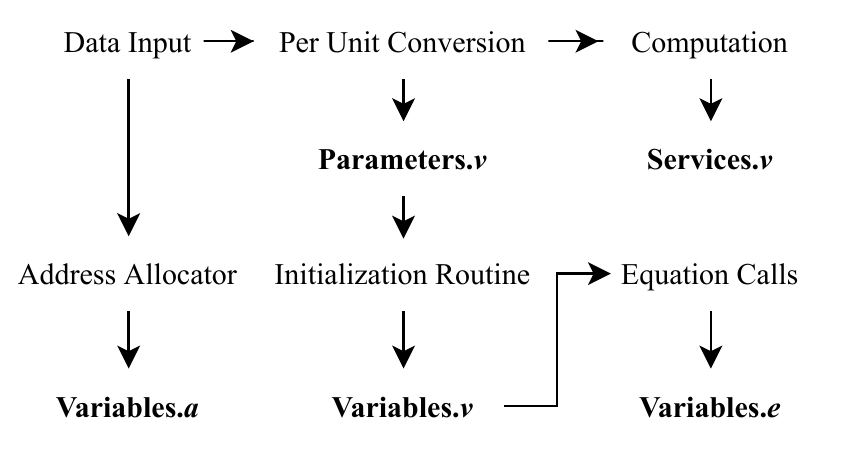}
\caption{Data flow paths for setting up and the numerical storage.}
\label{fig:data_path_numerical}
\end{figure}

Numerical arrays are updated at different phases in simulations, as outlined in \figref{fig:data_path_numerical}.
Parameter values are set after loading the data file and converting it to per unit under the system base. 
Variable addresses are allocated after loading the test system, values set by initialization calls, and equation values updated by equation calls. 
Service values are updated in multiple phases --- some are computed when accessed for the first time, and others are computed after parameters are set.
Discrete flags are updated before or after equation updates, depending on the discrete type.

\begin{table}[]
\caption{Modeling elements and their numerical attributes}
\label{tab:types_numerical_attributes}
\centering
	\begin{tabular}{@{}lcccc@{}}
		\toprule
		& Value (\textit{v}) & Address (\textit{a}) & Equation (\textit{e}) & Flags \\ \midrule
		Parameter & \checkmark   &      &    & \\
		Variable  & \checkmark   & \checkmark    & \checkmark  & \\
		Discrete  &     &      &    & \checkmark \\
		Service   & \checkmark   &      &    & \\
		\bottomrule
	\end{tabular}
\end{table}

\subsection{Variable Initialization}
Variable initialization routine sets variable initial values before a routine starts. 
It includes setting the starting point for power flow and initializing the rest of the variables for dynamic routines. 
Although power flow initialization is simple, there could be value conflicts depending on the input data format. 
For example, default initial bus voltages are set by buses and overwritten by PV-generators. 
The library uses an additional flag to indicate if the values from one model overwrite the shared variables at the end. 

Variable initialization for dynamics is mathematically a root-finding problem for the DAE system with all derivatives zeroed out. Two approaches can be used: sequential or iterative. Variables with an explicit solution can be initialized sequentially, while those without must be solved iteratively.

The library provides three entry points for initialization. First, an explicit-form equation can be specified for each variable if it can be initialized sequentially. A common technique is to set the initialization equation for a service that calculates from other services.
Second, an optional, implicit-form equation with an initial value can be specified for each variable. All implicit equations will be gathered and solved iteratively from the given initial value. 
Third, a placeholder function is available if one decides to write numerical code.
For best practice, sequential initialization should be used whenever possible. For convergence consideration, the initial values for the iterative initializers need to be carefully selected. 

\subsection{Numerical Equation Evaluation}
After loading a test case and counting the total number of variables, four numerical arrays are created to hold all variables and equations.
Each variable in a model receives an array of addresses indexing into the corresponding DAE array.
The same addresses can be used to access the corresponding numerical values of variables and equations. 

It is worth noting that the power system data structure introduces external variables for one model to link to another.
As shown in \autoref{lst:shunt_model}, the Shunt model creates two external algebraic variables, \lstinline{a} and \lstinline{v}, for linking to Bus devices with the indices given by \lstinline{bus}.
Variable addresses of the linked Bus devices will be assigned to Shunt so that Shunt has access to the Bus phase angles and voltages.

Memory copying of arrays imposes a significant overhead in numerical simulation. As a solution, all internal variables are assigned contiguous addresses so that a no-copy array view can be stored locally in each model. 
External variables are not guaranteed to link to contiguous devices, so their variables and equations are stored in local arrays and merged into the DAE arrays after evaluation.
Although this implementation is specific to \textit{NumPy}, the general rule applies to avoid memory copying, especially in computation-intense programs. 
However, one needs to realize the downside of this approach --- it rules out the possibility of parallelizing equation updates across models.
Since parallel equation updates are difficult in Python due to global interpreter lock (GIL), this shared-memory sequential evaluation approach will give the best performance.

Steps to update equations for each model are as follows.

\begin{figure}[!t]
\centering
\includegraphics[width=\columnwidth]{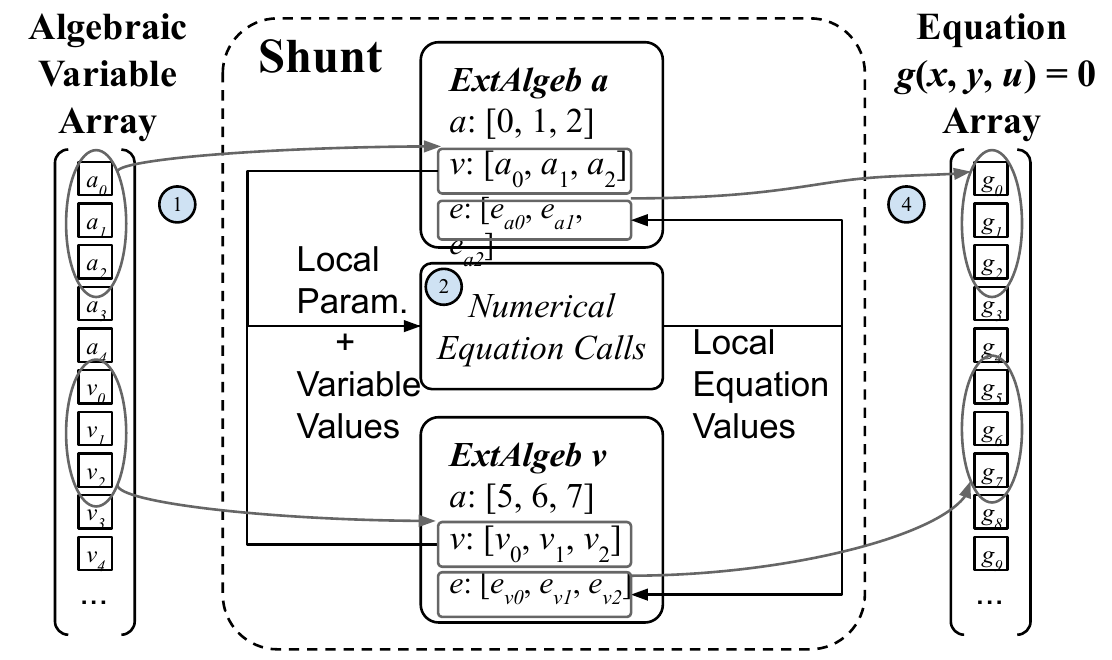}
\caption{Illustration of the equation update procedure (without Step 3).}
\label{fig:equation_update}
\end{figure}

\begin{enumerate}
    \item Copy external variables from DAE arrays to model.
    \item Call generated numerical functions using local values as inputs and store the outputs locally.
    \item Update equation values for equation-dependent limiters such as anti-windup limiters.
    \item Merge local external equation values to DAE equations. 
\end{enumerate}

This procedure (without step 3) is illustrated in \figref{fig:equation_update}. Note that step 3 is needed for models with equation-dependent limiters. Anti-windup limiters, for example, check the equation values to update the limiter status. Step 3 updates limiter status and sets the differential equation values to zero for the binding anti-windup limiters.

\subsection{Incremental Jacobian Building}
Building Jacobian matrices involve steps to fill in sparse Jacobian matrices incrementally and efficiently. 
It is especially relevant for implicit numerical integration routine since Jacobian updates take up the most overhead. 
This subsection discusses how the Jacobian indexing is done with the local variable indices (from the symbolic layer) and variable addresses (assigned in the numeric layer).

It is worth noting the difference between the local variable indices and the assigned variable addresses. 
A local variable index is a scalar number based on the sequence of declaration and is independent of test cases. 
Variable addresses are assigned as arrays after loading a specific test case. 
Local indices are used to look up corresponding addresses in order to determine the positions of the values.

For a generic triplet (\lstinline{row}, \lstinline{column}, \lstinline{value(*args)}) where \lstinline{row} and \lstinline{columns} are two scalars for the local indices, and \lstinline{value(*args)} is the numerical function for the Jacobian value with \lstinline{args} being a list of local values.
Recall that \lstinline{value} is the derivative of the \lstinline{row}-th equation with respect to the \lstinline{column}-th variable.
Jacobian values, which have the same length as the row and column addresses, should be summed at the positions defined by the case-specific addresses for the \lstinline{row}-th equation and the \lstinline{col}-th variables. 

\begin{figure}[!t]
\centering
\includegraphics[scale=0.8]{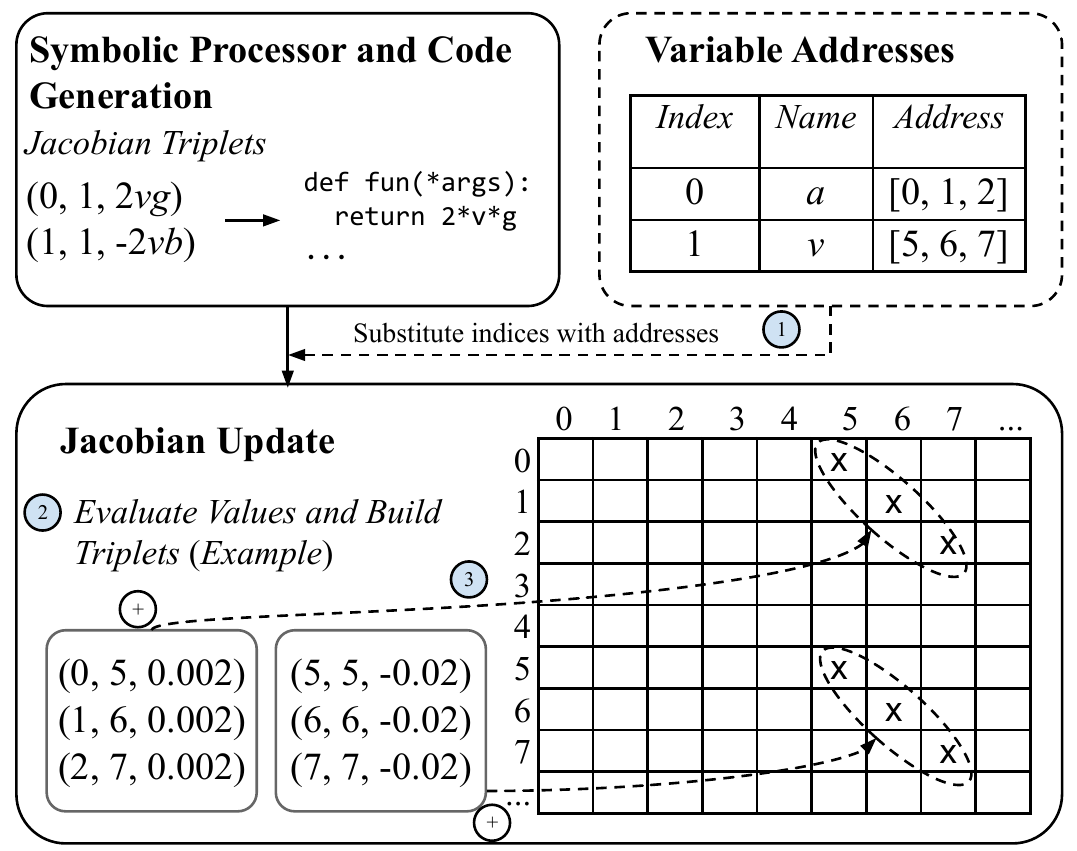}
\caption{Illustration of the Jacobian update procedure.}
\label{fig:jacobian_upate}
\end{figure}

\figref{fig:jacobian_upate} illustrates the process with three \lstinline{Shunt} devices as an example. 
There are two Jacobian triplets from the symbolic layer to be placed at local indices $(0, 1)$ and $(1, 1)$. 
In the numeric layer, the zeroth variable \lstinline{a} is assigned addresses $[0, 1, 2]$ and the first variable \lstinline{v} is assigned addressee $[5, 6, 7]$. Evaluate the numerical function $2vg$ to obtain the Jacobian elements, for example, $[0.002, 0.002, 0.002]$.
Next, these elements will be summed up at positions with the row number equal to the addresses of \lstinline{a} ($[0, 1, 2]$) and the column number equal to the address of \lstinline{v} ($[5, 6, 7]$). 
Repeat the process until all elements from all models are added.

For performance consideration, the library implements a two-step process that builds the sparsity pattern for one time and then fills in the values repeatedly. 
It is known that incrementally building sparse matrices can be time-consuming if repeated memory allocation is needed.
By using the addresses of elements, zero-filled sparsity pattern matrices can be constructed. 
The memory for the non-zero elements is pre-allocated, and in-place modifications can apply.
This technique is especially relevant for high-level languages without direct memory access. 
\section{Case Studies}
\label{sec:case-studies}
For verification and demonstration, this section presents a model implementation, power flow calculation, time-domain simulation, and eigenvalue analysis. 
The implementation of turbine governor model TGOV1 is demonstrated with source code developed in the proposed library.
Next, power flow results are reported with their time breakdown analyzed.
Further, time-domain simulation and eigenvalue analysis are verified against DSATools 19.0.

All subsequent studies are performed in CPython 3.7.7 with ANDES 1.0.3, SymPy 1.5.1, NumPy 1.18.4, and CVXOPT 1.2.5 on an AMD Ryzen 7 2700X CPU running Debian 10. In addition, a custom C-based routine is used for fast in-place sparse matrix addition. 

\subsection{Example Model: TGOV1}
\label{sec:tgov1_example}
\begin{figure}[!t]
\centering
\includegraphics[width=\columnwidth]{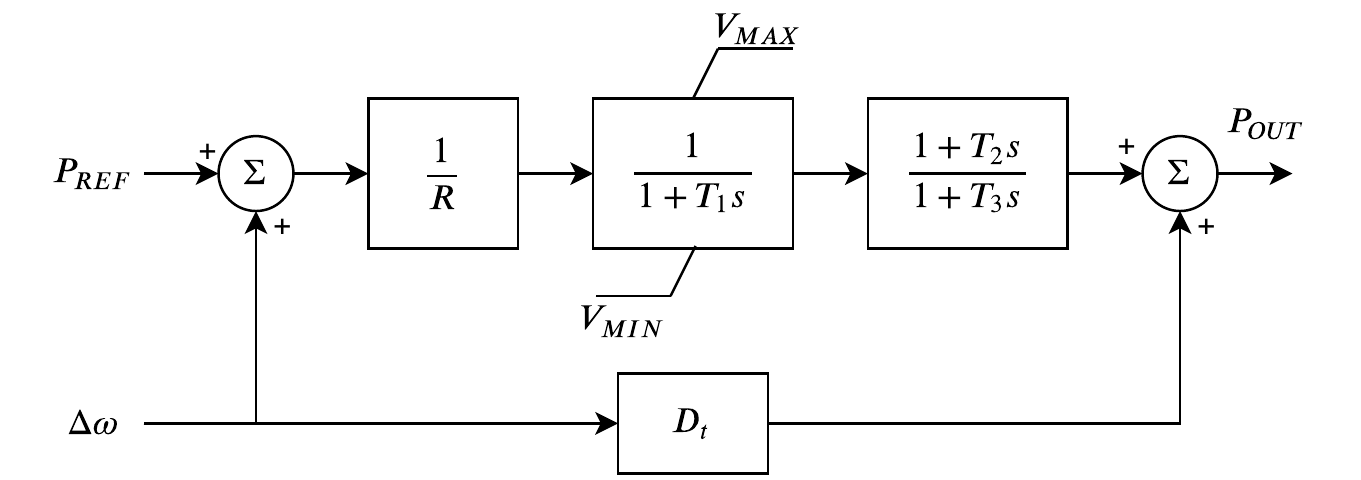}
\caption{The control diagram of TGOV1 turbine governor.}
\label{fig:tgov1_diagram}
\end{figure}

The TGOV1 turbine governor model \cite{Powerworld} (shown in \figref{fig:tgov1_diagram}) is used as a practical example with sufficient complexity to demonstrate the proposed work. 
This model is composed of a lead-lag transfer function and a first-order lag transfer function with an anti-windup limiter.
The corresponding differential equations and algebraic equations are given in \eqref{eq:tgov1_diff} and \eqref{eq:tgov1_algeb}.

\begin{equation}
    \left[
    \begin{matrix}
    \dot{x}_{LG} \\
    \dot{x}_{LL}
    \end{matrix}
    \right]
    =
    \left[
    \begin{matrix}z_{i,lim}^{LG} \left(P_{d} - x_{LG}\right) / {T_1}
    \\
    \left(x_{LG} - x_{LL}\right) / T_3
    \end{matrix}
    \right]
    \label{eq:tgov1_diff}
\end{equation}
\begin{equation}
    \left[
    \begin{matrix}
    0 \\
    0 \\
    0 \\
    0 \\
    0 \\
    0
    \end{matrix}
    \right]
    =
    \left[
    \begin{matrix}
    (1 - \omega) - \omega_{d} \\
    R \times \tau_{m0} - P_{ref} \\ 
    \left(P_{ref} + \omega_{d}\right)/R - P_{d}\\
    D_{t} \omega_{d} + y_{LL}  - P_{OUT}\\
    \frac{T_2}{T_3} \left(x_{LG} - x_{LL}\right) + x_{LL} - y_{LL}\\
    u \left(P_{OUT} - \tau_{m0}\right)
    \end{matrix}
    \right]
    \label{eq:tgov1_algeb}
\end{equation}
where \textit{LG} and \textit{LL} denote the lag block and the lead-lag block, $\dot{x}_{LG}$ and $\dot{x}_{LL}$ are the internal states, $y_{LL}$ is the lead-lag output, $\omega$ the generator speed, $\omega_d$ the generator under-speed, $P_d$ the droop output, $\tau_{m0}$ the steady-state torque input, and $P_{OUT}$ the turbine output that will be summed at the generator.
\begin{lstlisting}[
float,
floatplacement=H,
caption={Implementation of the TGOV1 model.},
captionpos=b,
label=lst:tgov1_code
]
def __init__(self):
  # 1. Declare parameters from case file inputs.
  self.R = NumParam(info='Turbine governor droop',
                    non_zero=True, ipower=True)
  # Other parameters are omitted to conserve space.
  
  # 2. Declare external variables from generators.
  self.omega = ExtState(src='omega', model='SynGen',
                 indexer=self.syn)
  self.tm = ExtAlgeb(src='tm', indexer=self.syn, 
              model='SynGen', e_str='u*(pout-tm0)')

  # 3. Declare services for temporary values.
  self.G = ConstService(e_str='u/R')
  self.tm0 = ExtService(src='tm',
               model='SynGen', indexer=self.syn)

  # 4. Declare variables and equations.
  self.pref = Algeb(v_str='tm0*R',
                e_str='tm0*R-pref')
  self.wd = Algeb(e_str='(1-omega)-wd')
  self.pd = Algeb(v_str='tm0',
              e_str='G*(wd+pref)-pd')
  self.LG_y = State(v_str='pd',
                e_str='LG_lim_zi*(pd-LG_y)/T1')
  self.LG_lim = AntiWindup(u=self.LG_y,
                  lower=self.VMIN,
                  upper=self.VMAX)
  self.LL_x = State(v_str='LG_y',
                e_str='(LG_y-LL_x)/T3')
  self.LL_y = Algeb(v_str='LG_y',
                e_str='T2/T3*(LG_y-LL_x)+LL_x-LL_y')
  self.pout = Algeb(v_str='tm0',
                e_str='(LL_y+Dt*wd)-pout')
\end{lstlisting}

An implementation of the TGOV1 model using descriptive equations is given in \autoref{lst:tgov1_code}. 
It consists of four types of declarations: parameters, external variables, initial external values, and internal variables and equations. 
Parameters are declared with special properties for data consistency and per-unit conversion.
For example, Line 4 specifies that the droop parameter $R$ must be non-zero and is an inverse-of-power per-unit quantity in device base MVA.
External variable $\omega$ is retrieved for calculation and $\tau_m$ for power feedback to generators. 
Note that the equation associated with $\tau_m$ replaces the steady-state constant torque $\tau_{m0}$ with the turbine output $P_{OUT}$.
The initial value of the mechanical torque is retrieved for variable initialization. 
Finally, differential and algebraic variables are declared, followed by the mathematical equations in \eqref{eq:tgov1_diff}-\eqref{eq:tgov1_algeb} written in a descriptive format, making it convenient to understand and troubleshoot. 

Alternatively, modeling blocks can be used to model part of TGOV1 directly from the transfer function diagram.
That is, lines 21-32 in \autoref{lst:tgov1_code} can be simplified into \autoref{lst:tgov1_simplification}, which is highly readable and similar to using a visual modeling tool in a scripting manner.
Note that variable \lstinline{pd} have been replaced with \lstinline{GA_y} in \autoref{lst:tgov1_simplification}, but the rest remain the same.
Modeling users can readily utilize blocks such as \lstinline{Gain}, \lstinline{LagAntiWindup} and \lstinline{LeadLag} without having to reimplement the underlying standard equations. 
\begin{lstlisting}[
float,
floatplacement=H,
caption={Block implementation of the three transfer functions in TGOV1.},
captionpos=b,
label=lst:tgov1_simplification]
  self.GA = Gain(u='wd+pref', K=self.G)
  self.LG = LagAntiWindup(u=self.GA_y, T=self.T1,
              K=1, lower=self.VMIN, upper=self.VMAX)
  self.LL = LeadLag(u=self.LG_y, T1=self.T2, 
                    T2=self.T3)
\end{lstlisting}
\subsection{Power Flow Calculation}
ANDES implements a Newton-Raphson method for power flow calculation as the first proof of concept.
Models for bus, PQ, PV, transmission line, and shunt are developed, and a full Newton-Raphson routine is implemented using the direct sparse linear solver KLU \footnote{KLU is not shipped with CVXOPT but is available through an add-on package \lstinline{cvxoptklu} (compilation required).}.
Unlike conventional power flow packages, the symbolically implemented line model does not implement an admittance matrix, although it is feasible to do so numerically.
Instead, vector computation of line injections into buses are used to maintain generality across models.

The power flow routine is benchmarked using test systems from MATPOWER 7.0. 
With the same settings and start points, ANDES is able to solve the cases listed in \Cref{tab:power-flow-time} and obtain identical results to that from MATPOWER. 
Note that the actual ANDES computation time is about 10\% shorter than these reported in the table since the profiler was turned on to obtain the time breakdown. 

The time breakdown exposes some interesting facts. 
Updating the numerical equations and solving the linear equations is relatively fast and takes up less than 30\% of the time. 
About half of the time is consumed for filling in Jacobian elements, even though an efficient C-based routine is used to modify values in place.
The Jacobian time, however, can be reduced by implementing a dishonest algorithm that avoids updating Jacobians at every iteration step.

\begin{table}[]
	\caption{Time breakdown (in seconds) for MATPOWER test cases}
	\label{tab:power-flow-time}
	\begin{tabular}{@{}llllll@{}}
		\toprule
		Name         & \begin{tabular}[c]{@{}l@{}}Total \\ Iterations\end{tabular} & \begin{tabular}[c]{@{}l@{}}Solve \\ Equations\end{tabular} & \begin{tabular}[c]{@{}l@{}}Update \\ Equations\end{tabular} & \begin{tabular}[c]{@{}l@{}}Build \\ Jacobians\end{tabular} & Total \\ \midrule
		300          & 6       & 0.002    & 0.002   & 0.008   & 0.016 \\
		1354pegase   & 6       & 0.006    & 0.002   & 0.020  & 0.038 \\
		2736sp       & 5       & 0.012    & 0.003   & 0.033  & 0.061 \\
		6515rte      & 5       & 0.036    & 0.006   & 0.091  & 0.189 \\
		9241pegase   & 7      & 0.092    & 0.013    & 0.232  & 0.421  \\
		ACTIVSg10k & 5        & 0.065    & 0.008    & 0.120  & 0.250  \\ 
		ACTIVSg25k & 8        & 0.281    & 0.028    & 0.526  & 0.982  \\ 
		\bottomrule
	\end{tabular}
\end{table}

\subsection{Time-Domain Numerical Integration}
To validate the numerical simulation results, ANDES is compared with the commercial package DSATools TSAT using Kundur's two area system, IEEE 14-bus system and Northeast Power Coordinating Council (NPCC) 140-bus system.
All PQ loads are converted to constant impedance after power flow calculation.
The implicit trapezoidal method is used with a fixed step size of $1/30$ second. 

\begin{figure}[!t]
\centering
\includegraphics[width=\columnwidth]{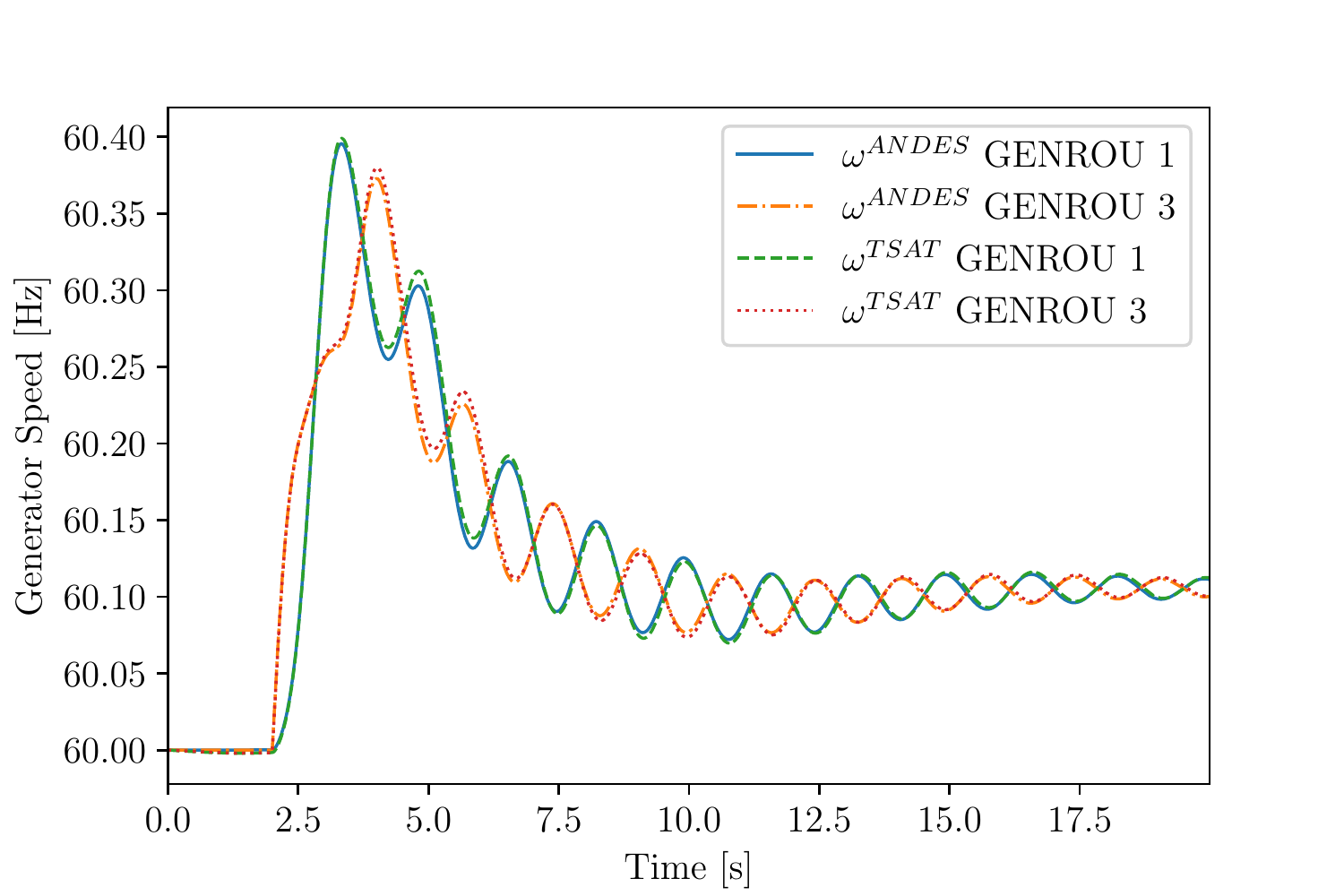}
\caption{The speed of generators on Buses 1 and 3}
\label{fig:kundur_speed}
\end{figure}

\begin{figure}[!t]
\centering
\includegraphics[width=\columnwidth]{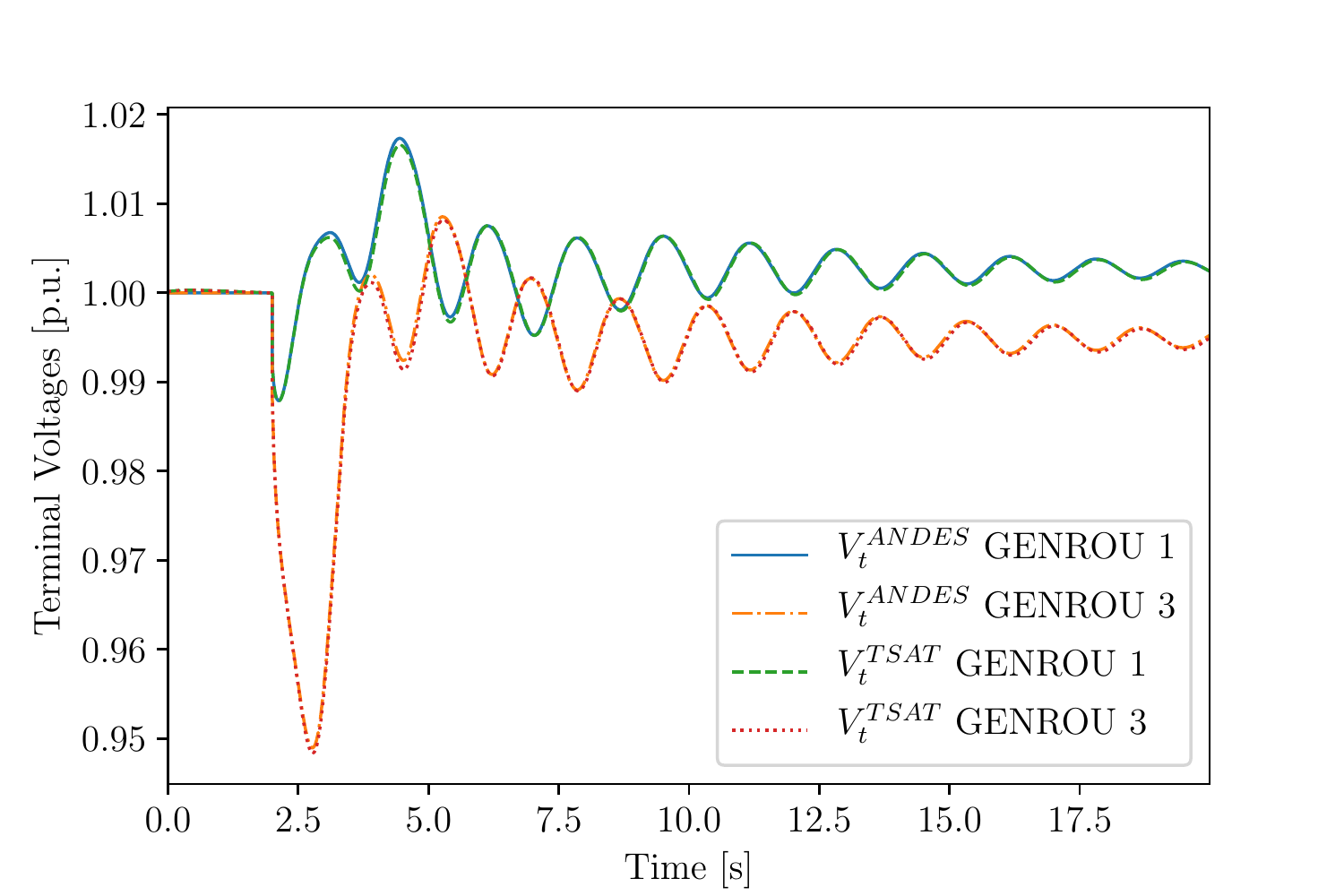}
\caption{Terminal voltages on Buses 1 and 3.}
\label{fig:kundur_voltage}
\end{figure}

\begin{figure}[!t]
\centering
\includegraphics[width=\columnwidth]{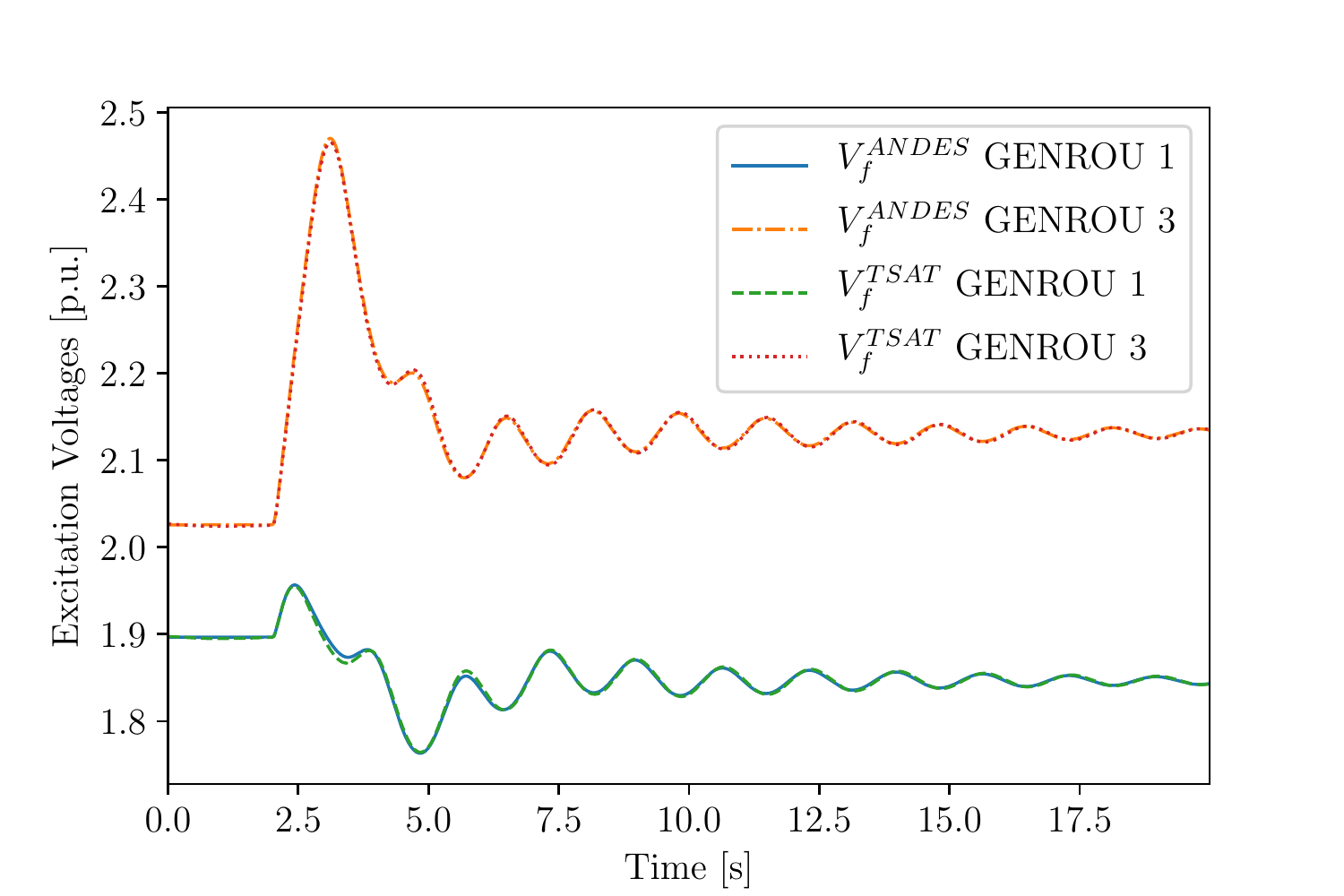}
\caption{Excitation voltages of generators on Buses 1 and 3.}
\label{fig:kundur_efd}
\end{figure}

The Kundur's system has four generators \cite{Kundur1994} in GENROU models \cite{Zhang2015a}, each with an EXDC2 exciter and a TGOV1 turbine governor. 
Parameters of the system are listed in the Appendix. 
At $t=2s$, one of the two lines between Bus 8 and Bus 9 is disconnected.
The simulation takes 1.2 seconds to complete.
Generator speed, terminal voltage, and excitation voltage following a line trip event are compared. 
Simulation results are depicted in \figref{fig:kundur_speed} - \figref{fig:kundur_efd}. 
Clearly, the proposed hybrid symbolic-numeric library achieves almost the same time-domain simulation results.

\begin{figure}[!t]
\centering
\includegraphics[width=\columnwidth]{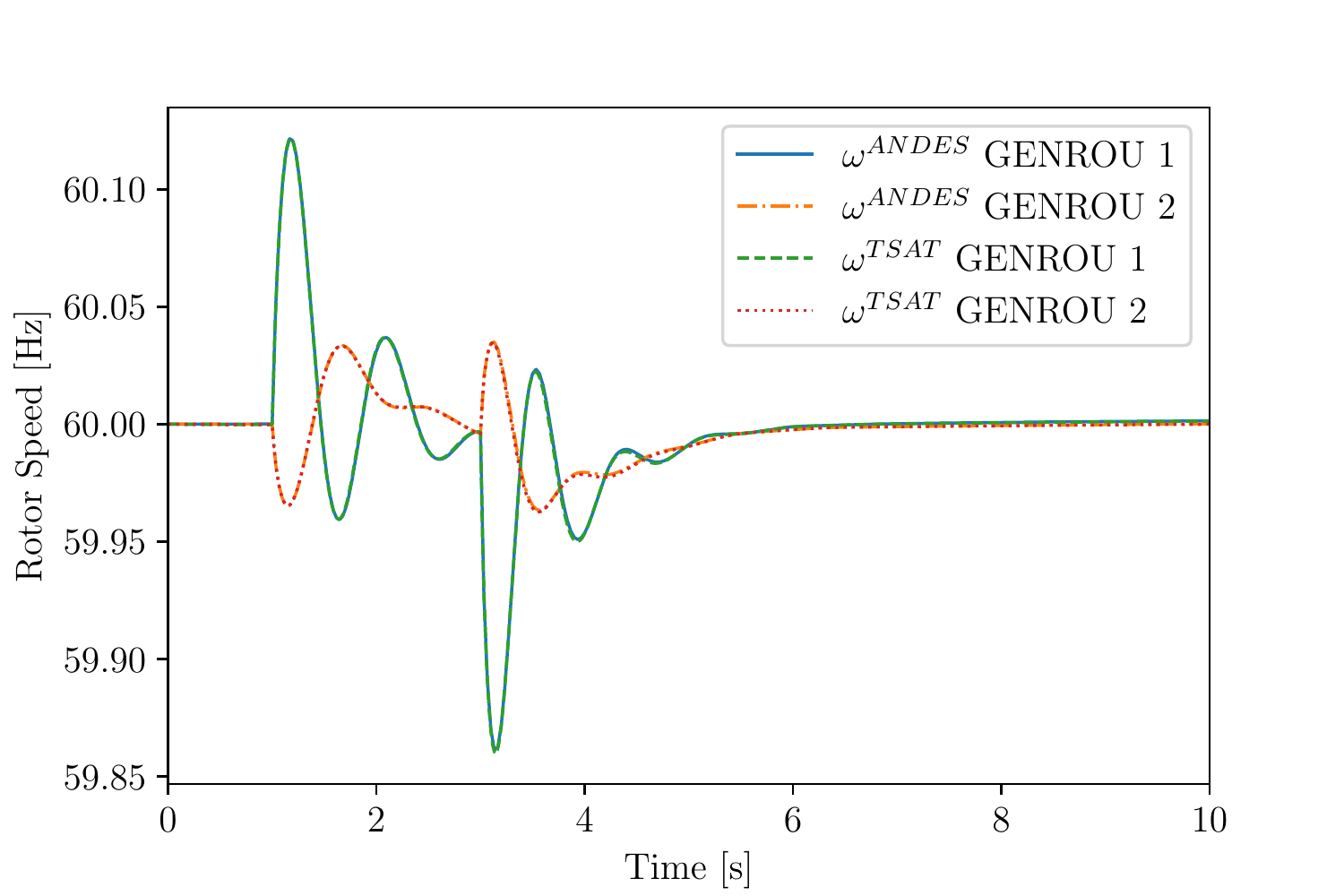}
\caption{IEEE 14-bus system rotor speed comparison.}
\label{fig:ieee14_omega}
\end{figure}

\begin{figure}[!t]
\centering
\includegraphics[width=\columnwidth]{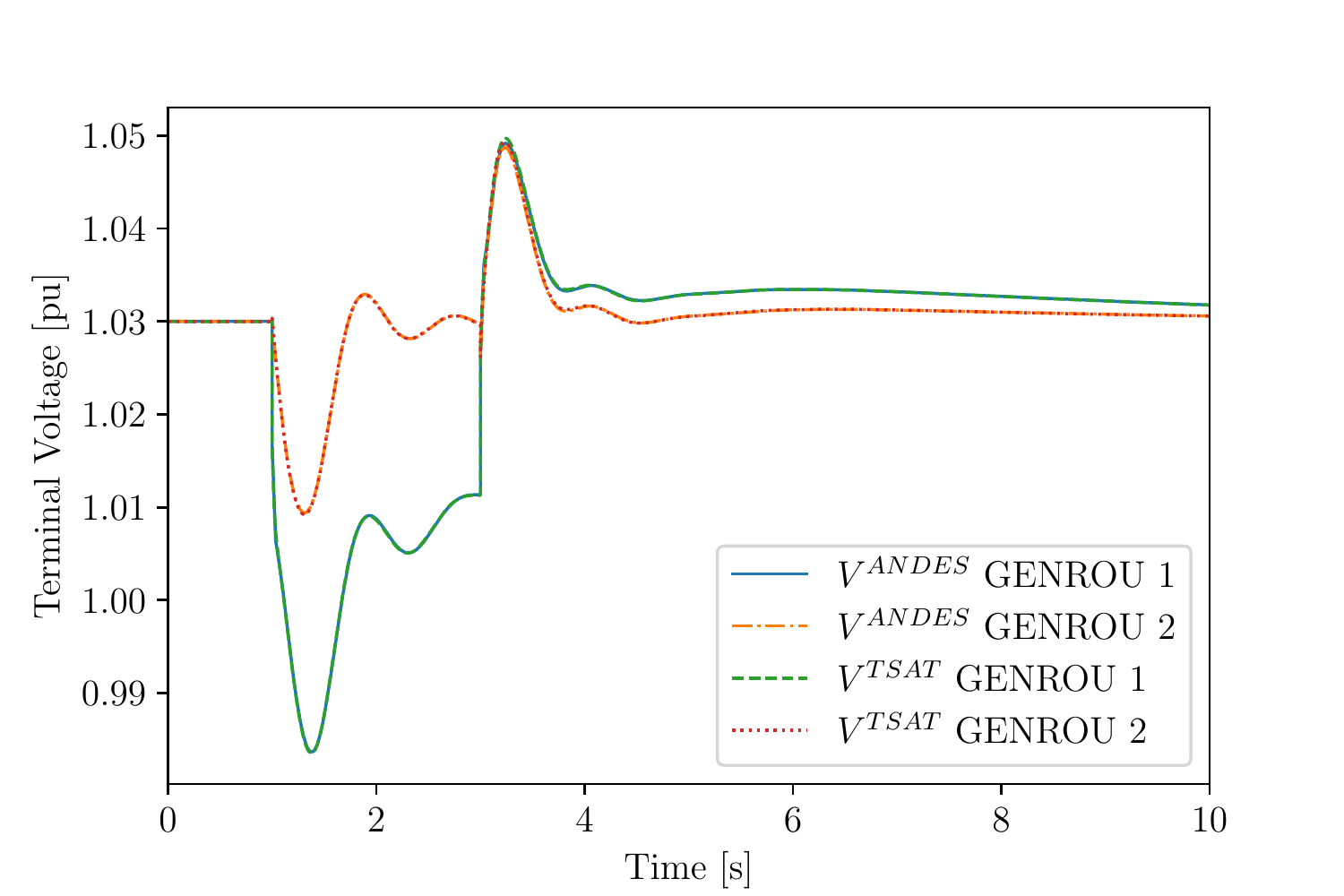}
\caption{IEEE 14-bus system voltage comparison.}
\label{fig:ieee14_v}
\end{figure}

The modified IEEE 14-bus system for validation uses a variety of models implemented in the hybrid symbolic-numeric framework.
These models include generator model GENROU, exciter models ESST3A and EXST1, turbine governor models TGOV1 and IEEEG1, and PSS models ST2CUT and IEEEST.
An extreme scenario that opens line 1-2 at 1 second and reconnects it after 2 seconds is used to trigger nonlinearity.
The simulation takes 4.1 seconds to complete.
Generator rotor speeds and terminal voltages in \figref{fig:ieee14_omega} and \figref{fig:ieee14_v} show perfect matches with TSAT.
The successful validation of ANDES using this system confirms the correct implementation of all the above models using the proposed framework.  

\begin{figure}[!t]
\centering
\includegraphics[width=\columnwidth]{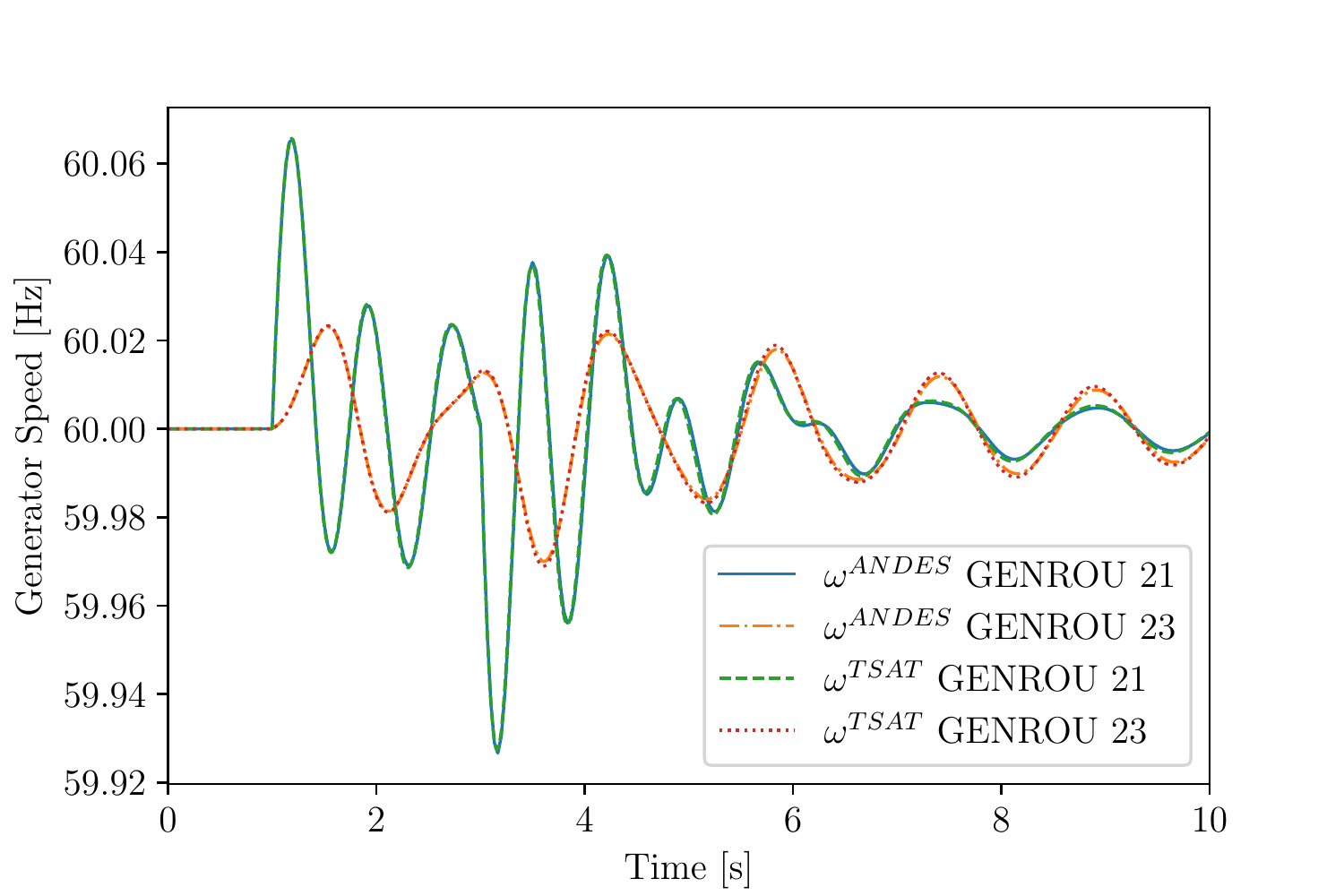}
\caption{NPCC 140-bus system rotor speed comparison.}
\label{fig:npcc_omega}
\end{figure}

\begin{figure}[!t]
\centering
\includegraphics[width=\columnwidth]{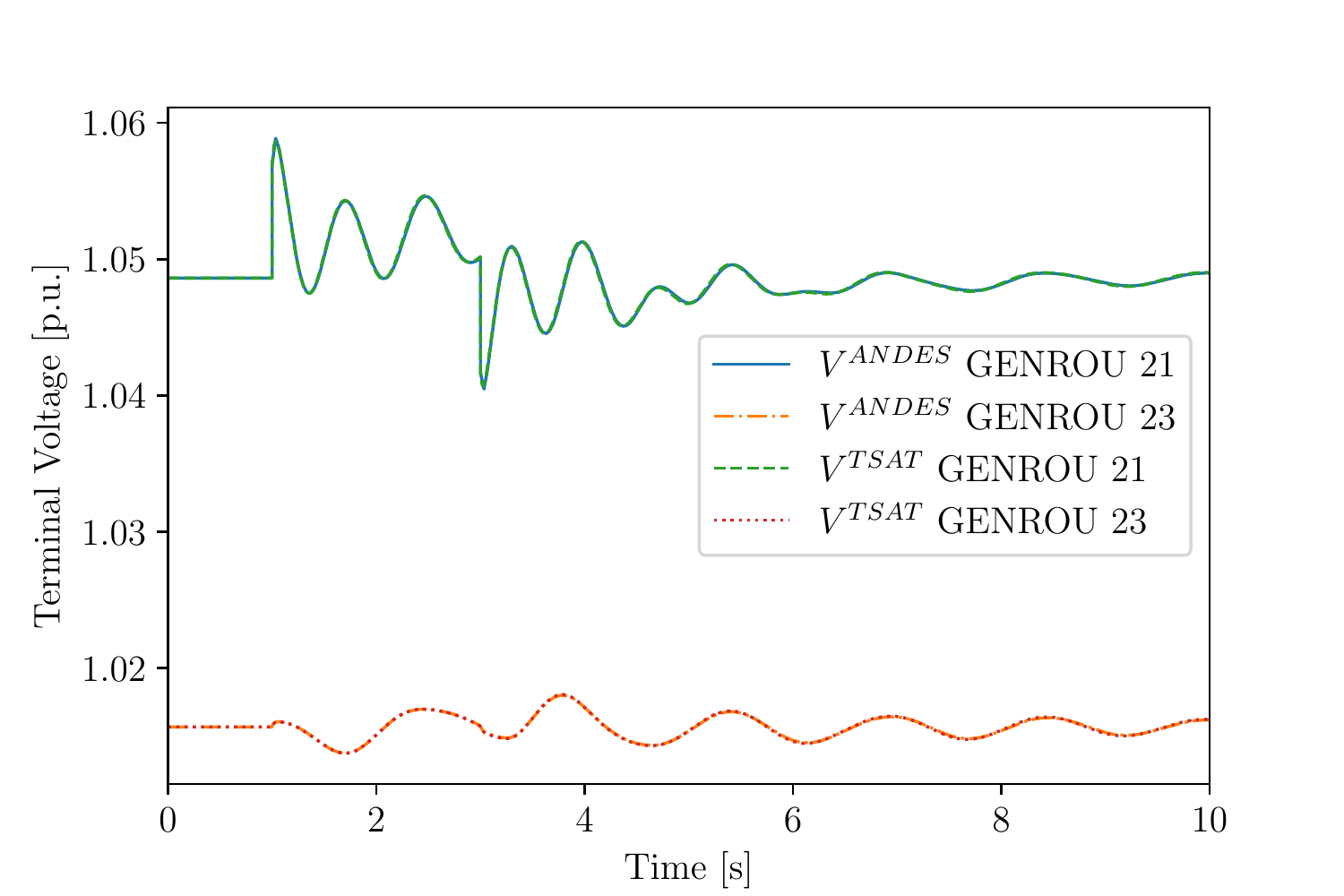}
\caption{NPCC 140-bus system voltage comparison.}
\label{fig:npcc_v}
\end{figure}

The NPCC 140-bus system (with generator models GENCLS and GENROU, exciter models IEEEX1 and turbine governor models TGOV1) is studied. 
The simulation takes 2.5 seconds to complete.
The rotor speed and voltage plots in \figref{fig:npcc_omega} and \figref{fig:npcc_v} also show perfect match.

It is also important to note that even commercial software does not always agree with each other, especially in large systems, due to factors such as unpublished implementation details and automatic parameter corrections.
Nevertheless, the discussed verification provides satisfactory results to prove the proposed concept using the above three test systems.

\subsection{Eigenvalue Analysis}
\begin{figure}[!t]
\centering
\includegraphics[width=\columnwidth]{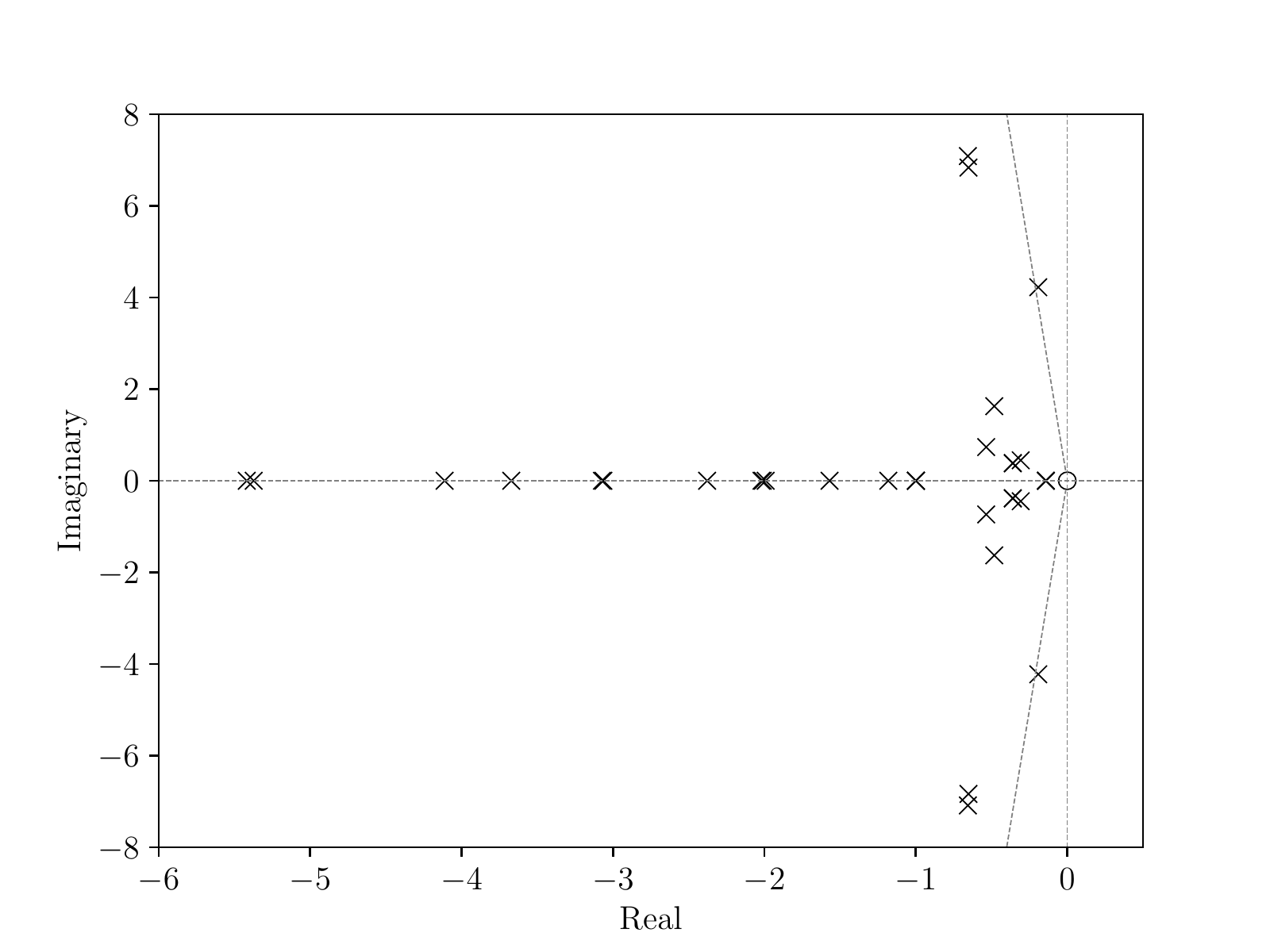}
\caption{Relevant eigenvalues in the \textit{S}-domain for Kundur's system.}
\label{fig:kundur_eigenvalue}
\end{figure}

Lastly, the numerical routine for eigenvalue analysis is developed by reusing existing eigenvalue programs.
Eigenvalues of the state matrix obtained after the time-domain initialization are plotted in \figref{fig:kundur_eigenvalue}. Two dotted lines in the figure are the loci with 5\% damping. 
Also, the first three eigenvalues ranked by damping ratio ($\zeta$) are compared between ANDES and DSATools SSAT in \Cref{tab:eigenvalues}.
The comparison shows that the numerical eigenvalue analysis routine in ANDES can obtain very close results to the commercial software SSAT.

\begin{table}
\centering
\caption{Eigenvalue results comparison for Kundur's system.}
\label{tab:eigenvalues}
\begin{tabular}{lllll}
\toprule
    & \multicolumn{2}{c}{ANDES}      & \multicolumn{2}{c}{SSAT}       \\
    \cmidrule(lr){2-3}
    \cmidrule(lr){4-5}
    & Eigenvalue      & $\zeta$ (\%) & Eigenvalue      & $\zeta$ (\%) \\
\midrule
\#1 & $-0.192\pm j4.225$ & 4.53         & $-0.192\pm j4.221$ & 4.55         \\
\#2 & $-0.656\pm j7.086$ & 9.22         & $-0.657\pm j7.083$ & 9.23         \\
\#3 & $-0.653\pm j6.834$ & 9.50         & $-0.653\pm j6.832$ & 9.51         \\
\bottomrule
\end{tabular}
\end{table}

\section{Conclusions}
\label{sec:conclusions}
In conclusion, this paper presents a hybrid symbolic-numeric library for DAE-based power system 
modeling and numerical simulation. 
This paper presented the design philosophy for a two-layer library that brings together the advantages of symbolic and numeric approaches. 
The symbolic layer is case-independent and handles descriptive modeling, symbolic processing, code generation, and automated documentation.
The numeric layer organizes the generated code for case-dependent initialization, equation update, and Jacobian update.
The simplicity of modeling using the proposed library is demonstrated with a TGOV1 turbine governor model. 
The library is verified for power flow calculation against MATPOWER, and the computation time is analyzed.
It is also verified for time-domain simulation using Kundur's system, IEEE 14-bus system, and NPCC system with a variety of dynamic models. 
The reference implementation in the ANDES library can obtain very close results for time-domain simulation and eigenvalue analysis to DSATools.

\section*{Acknowledgment}
The authors would like to thank Nicholas West for developing
the C-based routine for fast in-place sparse matrix addition.

\ifCLASSOPTIONcaptionsoff
  \newpage
\fi

\bibliographystyle{IEEEtran}
\bibliography{IEEEabrv,papers}

\begin{thebibliography}{10}
\providecommand{\url}[1]{#1}
\csname url@samestyle\endcsname
\providecommand{\newblock}{\relax}
\providecommand{\bibinfo}[2]{#2}
\providecommand{\BIBentrySTDinterwordspacing}{\spaceskip=0pt\relax}
\providecommand{\BIBentryALTinterwordstretchfactor}{4}
\providecommand{\BIBentryALTinterwordspacing}{\spaceskip=\fontdimen2\font plus
\BIBentryALTinterwordstretchfactor\fontdimen3\font minus
  \fontdimen4\font\relax}
\providecommand{\BIBforeignlanguage}[2]{{%
\expandafter\ifx\csname l@#1\endcsname\relax
\typeout{** WARNING: IEEEtran.bst: No hyphenation pattern has been}%
\typeout{** loaded for the language `#1'. Using the pattern for}%
\typeout{** the default language instead.}%
\else
\language=\csname l@#1\endcsname
\fi
#2}}
\providecommand{\BIBdecl}{\relax}
\BIBdecl

\bibitem{Jalili-Marandi2013}
\BIBentryALTinterwordspacing
V.~Jalili-Marandi, F.~J. Ayres, E.~Ghahremani, J.~Belanger, and V.~Lapointe,
  ``{A real-time dynamic simulation tool for transmission and distribution
  power systems},'' \emph{2013 IEEE Power {\&} Energy Society General Meeting},
  pp. 1--5, 2013. [Online]. Available:
  \url{http://ieeexplore.ieee.org/lpdocs/epic03/wrapper.htm?arnumber=6672734}
\BIBentrySTDinterwordspacing

\bibitem{chow1992toolbox}
J.~H. Chow and K.~W. Cheung, ``{A toolbox for power system dynamics and control
  engineering education and research},'' \emph{IEEE transactions on Power
  Systems}, vol.~7, no.~4, pp. 1559--1564, 1992.

\bibitem{zhou1996object}
E.~Zhou, ``Object-oriented programming, c++ and power system simulation,''
  \emph{IEEE Transactions on Power Systems}, vol.~11, no.~1, pp. 206--215,
  1996.

\bibitem{milano2005open}
F.~Milano, ``{An open source power system analysis toolbox},'' \emph{IEEE
  Transactions on Power systems}, vol.~20, no.~3, pp. 1199--1206, 2005.

\bibitem{Milano2013}
------, ``{A python-based software tool for power system analysis},'' in
  \emph{IEEE Power and Energy Society General Meeting}, 2013.

\bibitem{cole2011matdyn}
S.~Cole and R.~Belmans, ``{Matdyn, a new matlab-based toolbox for power system
  dynamic simulation},'' \emph{IEEE Transactions on Power systems}, vol.~26,
  no.~3, pp. 1129--1136, 2011.

\bibitem{Zhou2017}
\BIBentryALTinterwordspacing
M.~Zhou and Q.~Huang, ``{InterPSS: A New Generation Power System Simulation
  Engine},'' \emph{ArXiv e-prints}, 2017. [Online]. Available:
  \url{http://arxiv.org/abs/1711.10875}
\BIBentrySTDinterwordspacing

\bibitem{Top2016}
P.~Top, Y.~Qin, and L.~Min, ``{Integration of functional mock-up units into a
  dynamic power systems simulation tool},'' in \emph{IEEE Power and Energy
  Society General Meeting}, 2016, pp. 1--5.

\bibitem{Cui2018e}
H.~Cui and F.~Li, ``{ANDES : A Python-Based Cyber-Physical Power System
  Simulation Tool},'' in \emph{North American Power Symposium}, 2018, pp. 1--5.

\bibitem{cui2019cyber}
H.~Cui, F.~Li, and K.~Tomsovic, ``Cyber-physical system testbed for power
  system monitoring and wide-area control verification,'' \emph{IET Energy
  Systems Integration}, vol.~2, no.~1, pp. 32--39, 2019.

\bibitem{li2020large}
F.~Li, K.~Tomsovic, and H.~Cui, ``A large-scale testbed as a virtual power
  grid: For closed-loop controls in research and testing,'' \emph{IEEE Power
  and Energy Magazine}, vol.~18, no.~2, pp. 60--68, 2020.

\bibitem{Siemens}
Siemens, ``{PSS/E Graphical Model Builder}.''

\bibitem{PowerTech}
\BIBentryALTinterwordspacing
PowerTech, ``{DSATools}.'' [Online]. Available:
  \url{https://www.powertechlabs.com/dsatools-services}
\BIBentrySTDinterwordspacing

\bibitem{Milano2010}
F.~Milano, ``{Power system modelling and scripting},'' \emph{Springer}, 2010.

\bibitem{Baudette2018}
M.~Baudette, M.~Castro, T.~Rabuzin, J.~Lavenius, T.~Bogodorova, and
  L.~Vanfretti, ``{OpenIPSL: Open-Instance Power System Library — Update 1.5
  to “iTesla Power Systems Library (iPSL): A Modelica library for phasor
  time-domain simulations”},'' \emph{SoftwareX}, 2018.

\bibitem{Alvarado1988}
F.~L. Alvarado and Y.~Liu, ``{General Purpose Symbolic Simulation Tools for
  Electric Networks},'' \emph{IEEE Transactions on Power Systems}, vol.~3,
  no.~2, pp. 689--697, 1988.

\bibitem{Dzafic2002}
I.~D{\v{z}}afi{\'{c}}, F.~L. Alvarado, M.~Glavi{\'{c}}, and S.~Te{\v{s}}njak,
  ``{A Component Based Approach To Power System Applications Development},''
  \emph{14th PSCC (Power Syst. Computation Conf.)}, no. June, pp. 24--28, 2002.

\bibitem{Dzafic2004}
I.~Dzafic, M.~Glavic, and S.~Tesnjak, ``{A Component-Based Power System
  Model-Driven Architecture},'' \emph{IEEE Transactions on Power Systems},
  vol.~19, no.~4, pp. 2109--2110, 2004.

\bibitem{Alvarado1991}
F.~L. Alvarado, C.~A. Ca{\~{n}}izares, A.~Keyhani, and B.~Coates,
  ``{Instructional Use of Declarative Languages for the Study of Machine
  Transients},'' \emph{IEEE Power Engr. Review}, vol.~11, no.~2, p.~78, 1991.

\bibitem{Alvarado1996}
F.~L. Alvarado, C.~A. Ca{\~{n}}izares, and J.~Mahseredjian,
  ``{Symbolically-assisted power system simulation},'' \emph{International
  Journal of Electrical Power and Energy Systems}, vol.~18, no.~7, pp.
  405--408, 1996.

\bibitem{Gao2004}
W.~Gao, E.~V. Solodovnik, and R.~A. Dougal, ``{Symbolically aided model
  development for an induction machine in virtual test bed},'' \emph{IEEE
  Transactions on Energy Conversion}, vol.~19, no.~1, pp. 125--135, 2004.

\bibitem{Zimmerman2011b}
R.~D. Zimmerman, C.~E. Murillo-S{\'{a}}nchez, and R.~J. Thomas, ``{MATPOWER:
  Steady-state operations, planning, and analysis tools for power systems
  research and education},'' \emph{IEEE Trans. on Power Syst.}, 2011.

\bibitem{Meurer2017}
A.~Meurer, C.~P. Smith, M.~Paprocki, O.~{\v{C}}ert{\'{i}}k, S.~B. Kirpichev,
  M.~Rocklin, A.~T. Kumar, S.~Ivanov, J.~K. Moore, S.~Singh, T.~Rathnayake,
  S.~Vig, B.~E. Granger, R.~P. Muller, F.~Bonazzi, H.~Gupta, S.~Vats,
  F.~Johansson, F.~Pedregosa, M.~J. Curry, A.~R. Terrel, {\v{S}}.~Rou{\v{c}}ka,
  A.~Saboo, I.~Fernando, S.~Kulal, R.~Cimrman, and A.~Scopatz, ``{SymPy:
  Symbolic computing in python},'' \emph{PeerJ Computer Science}, 2017.

\bibitem{zhou_2012}
\BIBentryALTinterwordspacing
M.~Zhou, ``Interpss controller modeling language,'' 2012. [Online]. Available:
  \url{https://docs.google.com/document/d/1zvME4YBibCbEswVgS0PcqJdeA9ESMt9JAoBy7AGrr7c/preview}
\BIBentrySTDinterwordspacing

\bibitem{Cui2020rtd}
\BIBentryALTinterwordspacing
H.~Cui, ``{ANDES Documentation},'' 2020. [Online]. Available:
  \url{https://andes.readthedocs.io}
\BIBentrySTDinterwordspacing

\bibitem{Powerworld}
Powerworld, ``{Governor TGOV1 Model Reference}.''

\bibitem{Kundur1994}
P.~Kundur, \emph{{Power System Stability And Control}}.\hskip 1em plus 0.5em
  minus 0.4em\relax McGraw-Hill Inc., 1994.

\bibitem{Zhang2015a}
M.~Zhang, M.~Baudette, J.~Lavenius, S.~L{\o}vlund, and L.~Vanfretti,
  ``{Modelica Implementation and Software-to-Software Validation of Power
  System Component Models Commonly used by Nordic TSOs for Dynamic
  Simulations},'' in \emph{Proceedings of the 56th Conference on Simulation and
  Modelling (SIMS 56), October, 7-9, 2015, Link{\"{o}}ping University, Sweden},
  2015.

\end{thebibliography}

\newpage
\onecolumn
\appendix[Model Parameters for Kundur's Two Area System]
\label{apdx:parameter-kundur-two-area}

\setcounter{table}{0}
\renewcommand{\thetable}{A\arabic{table}}

\begin{longtable}{lrrrrr}
\caption{Bus Data}\label{tab:bus-data}\\
\toprule
{} &  idx &   Vn &     v0 &     a0 &  area \\
uid &      &      &        &        &       \\
\midrule
\endhead
\midrule
\multicolumn{6}{r}{{Continued on next page}} \\
\midrule
\endfoot

\bottomrule
\endlastfoot
0   &    1 &   20 &  1.000 &  0.570 &     1 \\
1   &    2 &   20 &  0.998 &  0.369 &     1 \\
2   &    3 &   20 &  0.963 &  0.185 &     2 \\
3   &    4 &   20 &  0.817 &  0.462 &     2 \\
4   &    5 &  230 &  0.979 &  0.480 &     1 \\
5   &    6 &  230 &  0.958 &  0.284 &     1 \\
6   &    7 &  230 &  0.936 &  0.127 &     1 \\
7   &    8 &  230 &  0.879 & -0.081 &     2 \\
8   &    9 &  230 &  0.891 &  0.094 &     2 \\
9   &   10 &  230 &  0.830 &  0.337 &     2 \\
\end{longtable}

\begin{longtable}{llrrrrrrr}
\caption{Line Data}\label{tab:line-data}\\
\toprule
{} &      idx &  bus1 &  bus2 &      r &      x &      b &  tap &  phi \\
uid &          &       &       &        &        &        &      &      \\
\midrule
\endhead
\midrule
\multicolumn{9}{r}{{Continued on next page}} \\
\midrule
\endfoot

\bottomrule
\endlastfoot
0   &   Line\_0 &     5 &     6 &  0.005 &  0.050 &  0.075 &    1 &    0 \\
1   &   Line\_1 &     5 &     6 &  0.005 &  0.050 &  0.075 &    1 &    0 \\
2   &   Line\_2 &     6 &     7 &  0.002 &  0.020 &  0.030 &    1 &    0 \\
3   &   Line\_3 &     6 &     7 &  0.002 &  0.020 &  0.030 &    1 &    0 \\
4   &   Line\_4 &     7 &     8 &  0.022 &  0.220 &  0.330 &    1 &    0 \\
5   &   Line\_5 &     7 &     8 &  0.022 &  0.220 &  0.330 &    1 &    0 \\
6   &   Line\_6 &     7 &     8 &  0.022 &  0.220 &  0.330 &    1 &    0 \\
7   &   Line\_7 &     8 &     9 &  0.002 &  0.020 &  0.030 &    1 &    0 \\
8   &   Line\_8 &     8 &     9 &  0.002 &  0.020 &  0.030 &    1 &    0 \\
9   &   Line\_9 &     9 &    10 &  0.005 &  0.050 &  0.075 &    1 &    0 \\
10  &  Line\_10 &     9 &    10 &  0.005 &  0.050 &  0.075 &    1 &    0 \\
11  &  Line\_11 &     1 &     5 &  0.001 &  0.012 &  0.000 &    1 &    0 \\
12  &  Line\_12 &     2 &     6 &  0.001 &  0.012 &  0.000 &    1 &    0 \\
13  &  Line\_13 &     3 &     9 &  0.001 &  0.012 &  0.000 &    1 &    0 \\
14  &  Line\_14 &     4 &    10 &  0.001 &  0.012 &  0.000 &    1 &    0 \\
\end{longtable}

\begin{longtable}{llrrr}
\caption{PQ Data}\label{tab:pq-data}\\
\toprule
{} &   idx &  bus &     p0 &     q0 \\
uid &       &      &        &        \\
\midrule
\endhead
\midrule
\multicolumn{5}{r}{{Continued on next page}} \\
\midrule
\endfoot

\bottomrule
\endlastfoot
0   &  PQ\_0 &    7 &  11.59 & -0.735 \\
1   &  PQ\_1 &    8 &  15.75 & -0.899 \\
\end{longtable}

\begin{longtable}{lrrrrrrr}
\caption{PV Data}\label{tab:pv-data}\\
\toprule
{} &  idx &  bus &  p0 &   q0 &  v0 &  ra &    xs \\
uid &      &      &     &      &     &     &       \\
\midrule
\endhead
\midrule
\multicolumn{8}{r}{{Continued on next page}} \\
\midrule
\endfoot

\bottomrule
\endlastfoot
0   &    2 &    2 &   7 &  3.0 &   1 &   0 &  0.25 \\
1   &    3 &    3 &   7 &  5.5 &   1 &   0 &  0.25 \\
2   &    4 &    4 &   7 & -1.0 &   1 &   0 &  0.25 \\
\end{longtable}

\begin{longtable}{lrrrrrrrr}
\caption{Slack Data}\label{tab:slack-data}\\
\toprule
{} &  idx &  bus &     p0 &     q0 &  v0 &  ra &    xs &    a0 \\
uid &      &      &        &        &     &     &       &       \\
\midrule
\endhead
\midrule
\multicolumn{9}{r}{{Continued on next page}} \\
\midrule
\endfoot

\bottomrule
\endlastfoot
0   &    1 &    1 &  7.459 &  1.436 &   1 &   0 &  0.25 &  0.57 \\
\end{longtable}

\begin{longtable}{lrrrrrrrrrrrrrrrr}
\caption{GENROU Data}\label{tab:genrou-data}\\
\toprule
{} &  idx &  bus &  gen &  D &      M &    xl &   xq &   xd &  xd1 &   xd2 &   xq1 &   xq2 &  Td10 &  Td20 &  Tq10 &  Tq20 \\
uid &      &      &      &    &        &       &      &      &      &       &       &       &       &       &       &       \\
\midrule
\endhead
\midrule
\multicolumn{17}{r}{{Continued on next page}} \\
\midrule
\endfoot

\bottomrule
\endlastfoot
0   &    1 &    1 &    1 &  0 &  13.00 &  0.06 &  1.7 &  1.8 &  0.3 &  0.25 &  0.55 &  0.25 &     8 &  0.03 &   0.4 &  0.05 \\
1   &    2 &    2 &    2 &  0 &  13.00 &  0.06 &  1.7 &  1.8 &  0.3 &  0.25 &  0.55 &  0.25 &     8 &  0.03 &   0.4 &  0.05 \\
2   &    3 &    3 &    3 &  0 &  12.35 &  0.06 &  1.7 &  1.8 &  0.3 &  0.25 &  0.55 &  0.25 &     8 &  0.03 &   0.4 &  0.05 \\
3   &    4 &    4 &    4 &  0 &  12.35 &  0.06 &  1.7 &  1.8 &  0.3 &  0.25 &  0.55 &  0.25 &     8 &  0.03 &   0.4 &  0.05 \\
\end{longtable}

\begin{longtable}{lrrrrrrrrrrrrr}
\caption{EXDC2 Data}\label{tab:exdc2-data}\\
\toprule
{} &  idx &  syn &    TR &    TA &  TC &  TB &    TE &    TF1 &    KF1 &  KA &  KE &  VRMAX &  VRMIN \\
uid &      &      &       &       &     &     &       &        &        &     &     &        &        \\
\midrule
\endhead
\midrule
\multicolumn{14}{r}{{Continued on next page}} \\
\midrule
\endfoot

\bottomrule
\endlastfoot
0   &    1 &    1 &  0.02 &  0.02 &   1 &   1 &  0.83 &  1.246 &  0.075 &  20 &   1 &    5.2 &  -4.16 \\
1   &    2 &    2 &  0.02 &  0.02 &   1 &   1 &  0.83 &  1.246 &  0.075 &  20 &   1 &    5.2 &  -4.16 \\
2   &    3 &    3 &  0.02 &  0.02 &   1 &   1 &  0.83 &  1.246 &  0.075 &  20 &   1 &    5.2 &  -4.16 \\
3   &    4 &    4 &  0.02 &  0.02 &   1 &   1 &  0.83 &  1.246 &  0.075 &  20 &   1 &    5.2 &  -4.16 \\
\end{longtable}

\begin{longtable}{lrrrrrrrrr}
\caption{TGOV1 Data}\label{tab:tgov1-data}\\
\toprule
{} &  idx &  syn &     R &  VMAX &  VMIN &    T1 &   T2 &  T3 &  Dt \\
uid &      &      &       &       &       &       &      &     &     \\
\midrule
\endhead
\midrule
\multicolumn{10}{r}{{Continued on next page}} \\
\midrule
\endfoot

\bottomrule
\endlastfoot
0   &    1 &    1 &  0.05 &    33 &   0.4 &  0.49 &  2.1 &   7 &   0 \\
1   &    2 &    2 &  0.05 &    33 &   0.4 &  0.49 &  2.1 &   7 &   0 \\
2   &    3 &    3 &  0.05 &    33 &   0.4 &  0.49 &  2.1 &   7 &   0 \\
3   &    4 &    4 &  0.05 &    33 &   0.4 &  0.49 &  2.1 &   7 &   0 \\
\end{longtable}

\end{document}